\documentclass{emulateapj}
\usepackage{epsfig}
\usepackage{color}
\usepackage{longtable}
\usepackage{natbib}
\newcommand{\etal}{{et~al.\ }}

\newcommand{\eg}{{\sl e.g.\/}}
\newcommand{\ie}{{\sl i.e.\/}}
\newcommand{\Akari}{\mbox{\sl AKARI\/}}
\newcommand{\GG}{\mbox{\sl GALEX\/}}
\newcommand{\HH}{\mbox{\sl Herschel\/}}
\newcommand{\IRAS}{\mbox{\sl IRAS\/}}
\newcommand{\ISO}{\mbox{\sl ISO\/}}
\renewcommand{\SS}{\mbox{\sl Spitzer\/}}

\newcommand{\Lsun}{\mbox{L$_\odot$}}
\newcommand{\Msun}{\mbox{M$_\odot$}}
\newcommand{\halpha}{\mbox{H$\alpha$}}

\slugcomment{Draft v. 2011 July 27}

\shorttitle{The Star Formation Reference Survey}
\shortauthors{Ashby et al.}

\begin{document}


\title{The Star Formation Reference Survey. \\
I. Survey Description and Basic Data}


\author
{
M.~L.~N.~Ashby\altaffilmark{1},
S.~Mahajan\altaffilmark{1,2},
H.~A.~Smith\altaffilmark{1},
S.~P.~Willner\altaffilmark{1},
G.~G.~Fazio\altaffilmark{1},
S.~Raychaudhury\altaffilmark{2},
A.~Zezas\altaffilmark{1,3}, 
P.~Barmby\altaffilmark{4},
P.~Bonfini\altaffilmark{3},
C.~Cao\altaffilmark{5},
E.~Gonz\'alez-Alfonso\altaffilmark{6},
D.~Ishihara\altaffilmark{7},
H.~Kaneda\altaffilmark{7},
V.~Lyttle\altaffilmark{4},
S.~Madden\altaffilmark{8},
C. Papovich\altaffilmark{9},
E.~Sturm\altaffilmark{10},
J.~Surace\altaffilmark{11},
H.~Wu\altaffilmark{12},
and
Y.-N.~Zhu\altaffilmark{12} \\
}


\altaffiltext{1}{Harvard-Smithsonian Center for Astrophysics, 60 Garden St., Cambridge, MA 02138, USA
[e-mail:  {\tt mashby@cfa.harvard.edu}]}

\altaffiltext{2}{School of Physics and Astronomy, University of Birmingham, Edgbaston, Birmingham B15 2TT, UK}

\altaffiltext{3}{University of Crete, Physics Department \& Institute of Theoretical and Computational Physics, 
71003 Heraklion, Crete, Greece}

\altaffiltext{4}{Department of Physics \& Astronomy, University of Western Ontario, London, ON N6A 3K7, Canada}

\altaffiltext{5}{School of Space Science and Physics, Shandong University at Weihai, Weihai, Shandong 264209, China}

\altaffiltext{6}{Universidad de Alcal\'a de Henares, Deptamento de F\'isica, 
Campus Universitario, E-28871 Alcal\'a de Henares, Madrid, Spain}

\altaffiltext{7}{Graduate School of Science, Nagoya University, Furo-cho, 
Chikusa-ku, Nagoya, Aichi 464-8602, Japan}

\altaffiltext{8}{Service d'Astrophysique, CEA/Saclay, l'Orme des Merisiers, 91191 Gif-sur-Yvette, France}

\altaffiltext{9}{Department of Physics, Texas A \& M University, College Station, TX 77843-4242}

\altaffiltext{10}{Max-Planck-Institute for Extraterrestrial Physics (MPE), 
Gießenbachstraße 1, D-85748 Garching, Germany}

\altaffiltext{11}{Spitzer Science Center, MS 220-6, California Institute of 
Technology, Jet Propulsion Laboratory, Pasadena, CA 91125, USA}

\altaffiltext{12}{Key Laboratory of Optical Astronomy, National Astronomical 
Observatories, Chinese Academy of Sciences, 20A Datun Road, 100012 Beijing, China}


\begin{abstract}
Star formation is arguably the most important physical process in the cosmos.
It is a fundamental driver of galaxy evolution and the ultimate source of most
of the energy emitted by galaxies.  A correct interpretation
of star formation rate (SFR) measures is therefore essential to our understanding
of galaxy formation and evolution.  Unfortunately, however, no single SFR estimator
is universally available or even applicable in all circumstances: the numerous
galaxies found in deep surveys are often too faint (or too distant) to 
yield significant
detections with most standard SFR measures, and until now there have been no
global, multi-band observations of nearby galaxies that span all the conditions
under which star-formation is taking place.  To address this need in a systematic
way, we have undertaken a multi-band survey of all types of star-forming galaxies 
in the local Universe.  This project, the Star Formation Reference Survey (SFRS), 
is based on a statistically valid sample of 369 nearby galaxies that span all 
existing combinations of dust temperature, SFR, and specific SFR.  Furthermore, 
because the SFRS is blind with respect to AGN fraction and environment it serves 
as a means to assess the influence of these factors on SFR.  Our
panchromatic global flux measurements (including \GG\ FUV+NUV, SDSS $ugriz$, 
2MASS $JHK_s$, \SS\ 3--8\,$\mu$m, and others) furnish uniform SFR measures and the context
in which their reliability can be assessed.  This paper describes the SFRS
survey strategy, defines the sample, and presents the multi-band photometry
collected to date.
\end{abstract}


\section{Introduction}

Star formation has been the most important single physical process since
recombination.  Not only have stars created most of the luminous energy in the 
Universe, they have 
produced the heavy elements needed for planets and life.  Star formation has naturally 
been the subject of vast numbers of studies, leading, for example, to our current 
understanding of the star formation history of the Universe (\eg, Lilly et al. 1996,
Hopkins \& Beacom 2006; Madau et al. 1998) and the discovery of the Schmidt Law 
(Schmidt 1959; updated by Kennicutt et al. 1998) relating the SFR in galaxies to 
the gas surface density.  

A general inadequacy of existing work is the lack of a
self-consistent treatment of SFR across the full electromagnetic spectrum:
ultraviolet (UV) continuum only samples the SFR in the absence of dust;
\halpha\ and [\ion{O}{2}] measure {\em ionizing} photons coming from
only the high-mass ($\ga$5\,\Msun) population.  Emission in the
polycyclic aromatic hydrocarbon (PAH) bands is taken as a SFR measure
(\eg, Madden \etal\ 2006; Wu \etal\ 2005) but may be
unreliable for low-luminosity (or low-metallicity) galaxies (Hogg \etal 2005)
or in the presence of a hard
radiation field.  Far-infrared dust re-radiation samples a broad range of star formation
($\sim$1--10\,\Msun\ for reasonable IMFs) but only becomes a precise
SFR measure in the optically-thick limit (Kennicutt 1998).  Nonthermal radio
emission comes from Type II supernova remnants (\eg, Rieke \etal 1980)
and therefore only samples higher-mass SF.

Star formation activity is by no means
evenly distributed throughout galaxies.  Instead, stars form within the 
densest regions of giant molecular clouds; this phenomenon manifests in the 
\IRAS\ bands as far-infrared radiation when the (hot) young stellar objects 
deeply embedded in their natal clouds illuminate the surrounding interstellar 
material (\eg, Parker 1991), which then reradiates that
energy at long wavelengths.  In a detailed case study of Milky Way GMC complexes 
(\eg, the California Nebula) Lada, Lombardi, \& Alves (2010) 
demonstrated a remarkably tight correlation between SFR and the total mass of 
{\sl dense} gas, lending further support to this view. 

There is considerable evidence in the literature that our understanding
of star formation depends on making consistent use of all available
wavebands.  For example, 
Kartaltepe \etal (2010) showed that without photometry at wavelengths
longer than 100\,$\mu$m, total far-IR luminosities (and thus SFRs) 
are typically underestimated by 0.2\,dex; in some cases the discrepancy 
can be much larger.  This can be interpreted as an inability to
adquately measure emission from a cold dust component (if present) when
such far-IR data are lacking.  
These issues are now
becoming better understood as a result of \Akari\ and \HH\ 
programs that reach deeper into the far-IR than could \IRAS, \ISO, or \SS. 
The situation remains complex, however.
In \HH-selected galaxies, dust
attenuation appears to strongly impact the UV detection fraction and
the relationship between UV and IR SFR indicators (Buat \etal 2010).
There are strong hints of systematic discrepancies between \Akari\ and
\IRAS\ photometry at $\sim$100\,$\mu$m (Jeong et al. 2007; Figs.~1 and 7 of 
Takeuchi et al 2010).
These unresolved issues can in principle be addressed with
thoughtful controls.  But there are also systematic effects intrinsic
to the galaxies themselves.  For example, there is evidence that the specific SFR 
(sSFR, the SFR per unit stellar mass) depends on stellar mass 
(\eg, Sobral \etal 2011, Elbaz \etal 2011).  For many years there have been suggestions that
dust temperature was dependent on luminosity (\eg, Sanders \etal 2003).  And
there is as yet no systematic treatment of the contribution of the older, 
quiescent stellar components to the SFR tracers most commonly used.

To better understand the complexities of star formation, a comprehensive treatment examining the
influence of all the major parameters --- stellar mass, dust mass and temperature,
and metallicity --- on SFR indicators is needed. 
The importance and timeliness of this subject is attested by 
numerous recent efforts to grapple with the nuances of SFR estimation.
Notable examples include the \SS\ Infrared Nearby Galaxies Sample 
(SINGS; Kennicutt \etal 2003; Calzetti \etal 2010), 
the Local Volume Legacy Survey (LVLS; Dale \etal 2009), the
\HH\ Reference Survey (HRS; Boselli \etal 2010), the Great Observatories All-sky
LIRG Survey (GOALS; Armus \etal 2009), and the Multi-Wavelength Extreme
Starburst Sample (MESS; Laag \etal 2010), among others.  
Each of these undertakings has been designed to attack specific aspects
of the star-formation phenomenon, but each also has limitations that prevent
it from providing a comprehensive picture of star formation in galaxies.  

This paper presents the Star Formation Reference Survey (SFRS), 
a sample of nearby galaxies having a unique capacity to describe star
formation under all conditions in which it occurs in the local
Universe.
Section~\ref{sec:selection} describes the selection criteria used to define 
the SFRS.  Section~\ref{sec:datasets} presents the SFRS
datasets obtained to date.  Section~\ref{sec:discussion} shows how 
the SFRS complements the above-mentioned projects, presents an
estimate of AGN prevalence and the ramifications 
for far-infrared SFR measures, and briefly describes 
SFRS-related observing campaigns now in progress.

\section{Sample Selection}
\label{sec:selection}

The complexities of the real Universe mean that an understanding
of global galaxy properties will inherently be statistical in nature.
Progress, even in the local Universe, requires uniform measurements
of the properties of a well-chosen, sufficiently large number of galaxies
that they sample star formation across {\sl the full range of galaxy
properties.}  In other words, having the proper study sample is critical 
to the success of this undertaking.

The SFRS selection criteria were defined objectively to guarantee
that the sample spans the full range of
properties exhibited by star-forming galaxies in the local Universe.
We began with the \IRAS\ 60\,\micron\ luminosity as an unbiased (but
perhaps not always correct!) star formation tracer by virtue of the breadth and
uniformity of \IRAS\
coverage and because of the proximity of the 60\,\micron\ band to the
SED peak associated with star formation.  Although the selection was based 
squarely on the 60\,\micron\ flux, this should not be taken as an
assertion that all issues of interpretation have been resolved.  
Some fraction of the 60\,\micron\ luminosity, at least in some galaxies, 
must arise from 
dust illuminated by the older stellar population.  Addressing this matter 
is one of the aims of the SFRS project and a primary motivation
for the assembly of the datasets described in Sec.~\ref{sec:datasets}.

The parent sample for the present study is the PSCz catalog
(Saunders \etal 2000), a full-sky database of 15,000 nearby
star-forming galaxies brighter than 0.6\,Jy at 60\,\micron.  Most
PSCz objects are closer than $z=0.2$.  The PSCz
is not biased toward relatively rare ultraluminous objects (unlike the 1\,Jy sample of
Kim, Veilleux, \& Sanders 2002) and is more representative than
more restricted samples (\eg, the \IRAS\ Bright
Galaxy Sample; $F_{60}>5.24$\,Jy; Sanders \etal 2003).  PSCz galaxy
luminosities range from $L(60\,\micron)=10^{7}$ to $10^{12}$\,\Lsun.

\begin{figure}
\includegraphics[width=80mm]{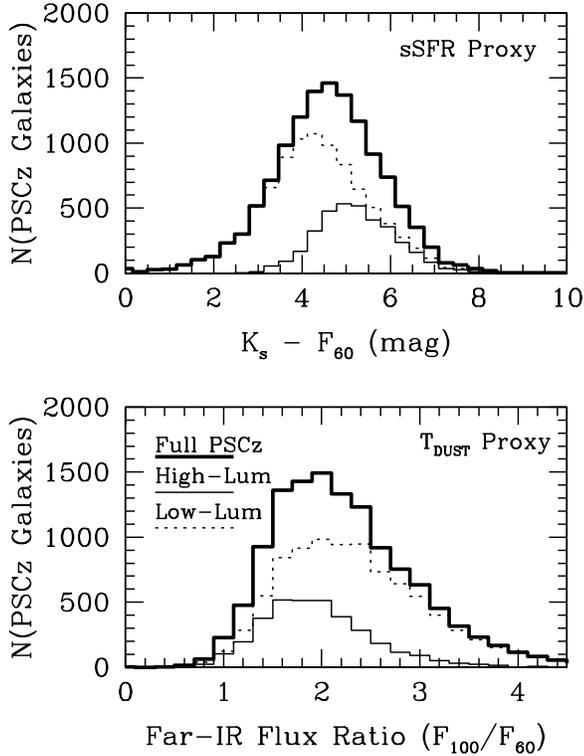}
\caption{
Histograms of sSFR and dust temperature proxies within the parent PSCz galaxy sample.
Thick lines show galaxy numbers for the full sample.  The PSCz has been divided into
two subsamples at $\nu L_\nu(60\,\micron) = 10^{9.5}$~\Lsun.  Thin solid lines indicate
the distributions for the high-luminosity subsample, and dashed lines indicate
the distributions for the low-luminosity subsample.
{\it Top panel:} Abscissa is the ratio of $K_s$ to 60\,\micron\ flux densities expressed as the
difference in AB magnitudes.  {\it Bottom panel:} Abscissa is the ratio of 100 to 60\,\micron\
flux densities ($F_\nu$).
}
\label{fig:tdust}
\end{figure}

Star-forming galaxies have a wide range of properties, but aside from
absolute luminosity, the specific SFR and the color temperature of far 
infrared dust emission are arguably the most important.  These parameters reflect 
the relative importance of present and past star formation and the densities of 
the regions where stars are forming.
High color temperature is usually taken to indicate that dust grains are close to the
newly-formed stars, \ie, it is a measure of active star formation.  A low
color temperature (high $F_{100}/F_{60}$ ratio), on the other hand, can
be interpreted as reflective of a situation in which the ambient UV field
from older stars in the galaxy disk is illuminating dust in the ISM.
Figure~\ref{fig:tdust} illustrates one way in which 
dust temperature and luminosity relate.  Nearly all `cool' galaxies 
(\ie, those with a far-IR flux density ratio $F_{100}/F_{60} \ge 2$)
reside in the low-luminosity regime. 
The situation is different for `warmer' ($F_{100}/F_{60} < 2$) galaxies, 
however -- this population is roughly evenly divided between high- and low-luminosity
sources.  To capture this behavior in a study sample, it is not 
adequate to make a simple luminosity cut: one must account for the dust
temperature or be at risk of under-representing the mode (warm or cool) 
at which star formation occurs for some luminosities.

\begin{figure}
\includegraphics[width=70mm]{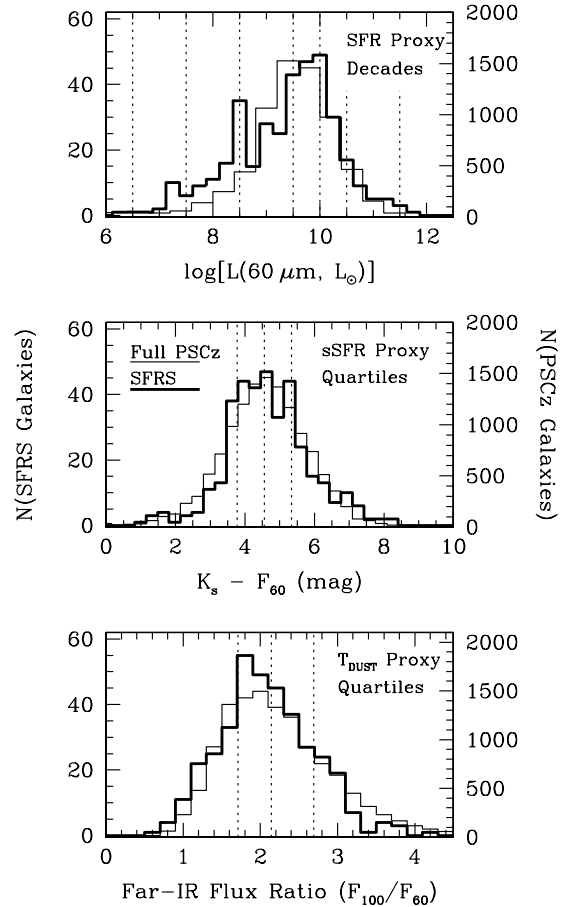}
\caption{
The distributions of the SFRS galaxies in each of the three parameter spaces
used to select the sample.  Thick histograms indicate the SFRS galaxies and
are referenced to the left-hand axes, while
thin histograms represent the larger PSCz sample from which the SFRS sample
is drawn, and are referenced to the right-hand axes.  The distributions are
very similar in all three panels.  The vertical dotted lines in
each panel indicate the boundaries defining the bins (Tab.~\ref{tab:bins}).
}
\label{fig:selection}
\end{figure}

\begin{deluxetable}{cccc}
\tablecolumns{3}
\tabletypesize{\scriptsize}
\tablecaption{Bin Boundaries Applied to Parent PSCz Sample\label{tab:bins}}
\tablewidth{0pt}
\tablehead{
\colhead{Parameter} & \colhead{Q1/Q2} & \colhead{Q2/Q3} & \colhead{Q3/Q4} \\
}
\startdata
sSFR Proxy ($K_s - F_{60}$) & 3.78 & 4.56 & 5.34 \\
$T_{DUST}$ Proxy ($F_{100}/F_{60}$) & 1.71 & 2.14 & 2.69 \\
\enddata
\tablecomments{The quartile boundaries applied to the sSFR and $T_{DUST}$ proxies
used to define the SFRS sample.  The third selection dimension, 60\,$\mu$m luminosity,
was binned in decades starting at log(L$_{60}$) = 10$^{6.5}$\,\Lsun\ with an
additional half-decade bin boundary at 10$^{10}$\,\Lsun, near the peak of the distribution
(Fig.~\ref{fig:selection}).
}
\end{deluxetable}

Figure~\ref{fig:tdust} also reveals a dichotomy in
the relationship of sSFR and luminosity.  While essentially all the low-sSFR
galaxies are low-luminosity objects, those with high sSFRs are a mix, with
both low- and high-luminosity galaxies present.  A selection on luminosity
alone is therefore extremely unlikely to include sources with the full range
of sSFRs that exist, and a high-luminosity selection will miss the low-sSFR
sources entirely.

A full sampling of star-forming galaxy properties requires that the SFRS be 
defined in a three-dimensional space spanning the {\sl full ranges} occupied
by PSCz galaxies in 60\,\micron\ luminosity, flux ratio $F_{60}$/$K_s$,
and far-IR flux density ratio $F_{100}/F_{60}$.
In this scheme 2MASS $K_s$
serves as a proxy for stellar mass, the far- to near-infrared flux
ratio measures specific SFR, while the far-IR flux density ratio
acts as a measure of dust temperature.  
This three-dimensional parameter space was binned by decade in 60\,\micron\ luminosity
and by quartiles in both the $K_s-60$\,\micron\ color and the $F_{100}/F_{60}$ flux density
ratio (Fig.~\ref{fig:selection}).  The bins are defined in Table~\ref{tab:bins}.
This ensured that the sample contains all existing combinations of high-
and low-sSFR with far-IR color {\em and} luminosity.  
The changeover from cool dust/low sSFR to warm dust/high sSFR occurs
around $L(60) = 10^{9.5}$ to $10^{10.5}$\,\Lsun, which is the most
heavily populated decade in luminosity.  
We therefore split this range
into two half-decades in $L(60)$ to sample this important regime more
finely.  The result is $4\times4$ bins in each of two colors and eight bins
in luminosity: 128 bins altogether.

After the bin boundaries were determined from the complete PSCz, 
a representative subsample was selected as follows.  First, only galaxies 
with measured positive redshifts were included.  Second, galaxies outside the 
Sloan Digital Sky Survey (SDSS)
and NRAO VLA Sky Survey (NVSS; Condon \etal 1998) of the northern Galactic cap were excluded.  
This constraint ensures 
that visible and radio detections, and hence radio-derived SFRs, would be available 
for all galaxies and simultaneously minimizes foreground
confusion because the Galactic plane is $>$20\degr\ distant from all sources.  
Out of the 15,000 PSCz galaxies, 2564 meet these restrictions.  
Third, the few nearest galaxies were eliminated in order to avoid 
time-consuming \SS\ mapping of objects with large angular diameters 
and the attendant uncertainties in calculating global measurements.  
The exact distance limit was made luminosity-dependent: 
at $L(60) < 10^{7.5}$\,\Lsun\ and $L(60) > 10^{10}$\,\Lsun, 
all galaxies were eligible because they were either sufficiently small or distant
that they could be imaged efficiently with \SS.  At intermediate luminosities,
galaxies with recession velocities $cz < 0.05 * L(60)^{1/2}$\,km\,s$^{-1}$ were excluded,
where $L(60)$ is expressed in units of \Lsun.  This criterion excluded 194 galaxies.
Because all these restrictions are based
strictly on galaxy positions in three-dimensional space, 
they in no way bias the sample, and
they ensure that low-luminosity sources are retained.

\begin{figure}
\includegraphics[width=75mm]{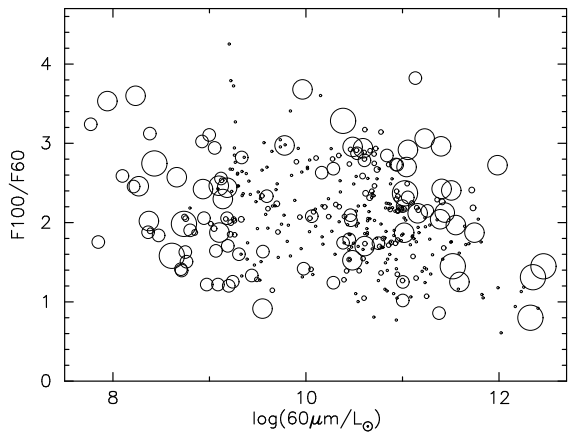}
\caption{
The three-dimensional SFRS galaxy distribution projected into the two-dimensional space
defined by 60\,\micron\ luminosity and far-IR flux density ratio.
The symbol size is inversely proportional to weight:
large symbols indicate relatively rare objects that occupy sparsely-populated
bins; they significantly enlarge the parameter space explored by our sample
and may be under-represented by programs not implementing selections similar
to the SFRS.
}
\label{fig:sample-1}
\end{figure}

\begin{figure}
\includegraphics[width=75mm]{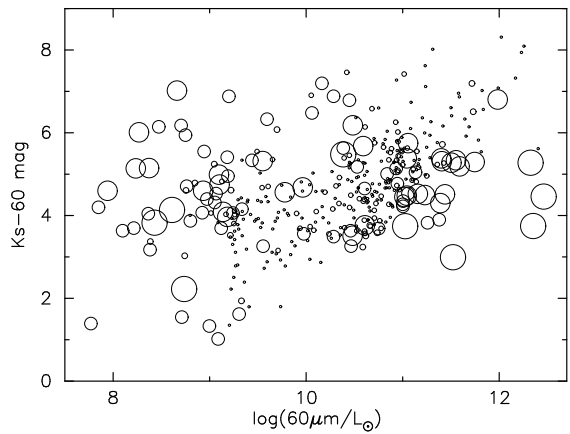}
\caption{
The three-dimensional SFRS galaxy distribution projected into the two-dimensional space
defined by 60\,\micron\ luminosity and near- to far-IR color (specific SFR proxy).
Symbols as in Fig.~\ref{fig:sample-1}.
}
\label{fig:sample-2}
\end{figure}

The final SFRS consists of a representative number of 
galaxies from each bin.  The number chosen from each
bin containing $N$ galaxies was $\sqrt N$, rounded up 
to a maximum of 10 galaxies.  The specific galaxies chosen were the brightest 
within each bin.  This selection sets the sample size at 369 galaxies and 
guarantees that the sample will be representative of the much larger PSCz, 
and by inference, representative of star forming galaxies generally.  
The distributions of the sample galaxies in the 
two-dimensional projections of the three-dimensional selection space 
are illustrated in Figs.~\ref{fig:sample-1} and \ref{fig:sample-2}.
The SFRS distributions are very similar in all three selection 
parameters (Fig.~\ref{fig:selection}) to those for the full PSCz.
The resulting sample is thus statistically well-defined,
\emph{and covers the entire range of star formation properties seen locally:
five decades in luminosity (and thus SFR); a factor of nearly
200 in specific SFR, and all masses, morphologies, and sizes.}

The relative prevalence of a galaxy of any type within
the sample is reflected in its weight: each
SFRS galaxy in a given selection bin is assigned a weight determined by 
the ratio of the total number of PSCz galaxies in that bin to the number
of sample galaxies drawn from that bin into the SFRS sample.  Relatively
rare galaxy types have low weights, while those from heavily populated
bins have large weights.  
The names, positions, individual weights, and \IRAS\ 60 and 100\,\micron\ fluxes used to
define the sample are given in Table~\ref{tab-basic}.  These weights project the
SFRS sample back to the parent PSCz population.  Additional weighting based on
the volume in which a galaxy would enter the PSCz would be needed to define
a true volume-limited sample.

Because of the rigid selection criteria, the well-known quasar 3C\,273 and the the blazar
OJ\,287 are members of the SFRS.  Each is the only galaxy in its bin (weight$\equiv$1).  
While we retain these objects to preserve the demographics of our selection, they can be 
ignored for studies relating purely to star formation.

\section{Basic Data}
\label{sec:datasets}

In addition to the 2MASS and \IRAS\ data used to define the selection criteria, 
many other resources are available.  These are summarized in Table~\ref{tab-surveys}
and described in detail beginning with Sec.~\ref{ssec:irac}.

\subsection{Distances}
\label{sec-velocities}

\tablenum{3}
\begin{deluxetable*}{ccc}
\tabletypesize{\scriptsize}
\tablecaption{Summary of Available Data for SFRS Galaxies\label{tab-surveys}}
\tablewidth{0pt}
\tablehead{
\colhead{Bandpass}     & \colhead{Observatory} & \colhead{Sample Coverage} \\
}
\startdata
1.4\,GHz               & VLA/NVSS                       & 100\% \\
12,25,60,100\,\micron\ & \IRAS\                         & 100\% \\
65,90,140,160\,\micron\ & \Akari\ FIR All-Sky Survey    & 95\% \\
24\,\micron             &  \SS/MIPS                     & 70\% \\
3.6, 4.5, 5.8, 8.0\,\micron\              & \SS/IRAC    & 100\% \\
$JHK_s$                 & 2MASS                         & 100\% \\
$ugriz$                 & SDSS                          & 100\%    \\
Optical Spectra         & SDSS (fiber)                  & 57\%  \\
H$\alpha$               & NAOC                          & 30\% (campaign ongoing)         \\
0.13--0.28\,\micron     & \GG                           & 90\% to date \\
\enddata
\end{deluxetable*}

A significant fraction of the SFRS galaxies are very 
nearby, and can have peculiar velocities comparable to 
their Hubble flow velocities.  
\citet{tully08} have recently accumulated 
redshift-independent distance measurements for nearby galaxies by using alternate 
methods including the Tully-Fisher relation \citep{tully77}, Cepheids \citep{freedman01}, 
the luminosity of stars at the tip of the red giant branch \citep{karachentsev04,karachentsev06}, 
and surface brightness fluctuations \citep{tonry01}.  These measurements yield {\em quality} 
distances for nearby ($v\,<\,3000$\,km s$^{-1}$) galaxies with distance modulus 
uncertainties $<0.1$ mag.  \citet{tully08} estimated the distances of additional 
galaxies based on their association with groups/clusters. In total, their 
catalog\footnote{The Extra-galactic Distance Database, http://edd.ifa.hawaii.edu/}
includes 3,529 distance measurements (Tully, 2010, private communication.) for galaxies
with $v\,<\,10,000$\,km s$^{-1}$, of which 127 are in the SFRS. 

For the 242 galaxies lacking quality distances, the heliocentric velocities in the PSCz catalog 
were converted to a corrected recession velocity taking into account
the velocity field of Virgo, the Great attractor and the Shapley supercluster,
following \citet{mould00}.  
All cataloged PSCz heliocentric velocities agree with those provided by 
the NASA Extra-galactic Database (NED)\footnote{http://ned.ipac.caltech.edu/}, 
within the uncertainties.  

\begin{figure}
\includegraphics[width=88mm]{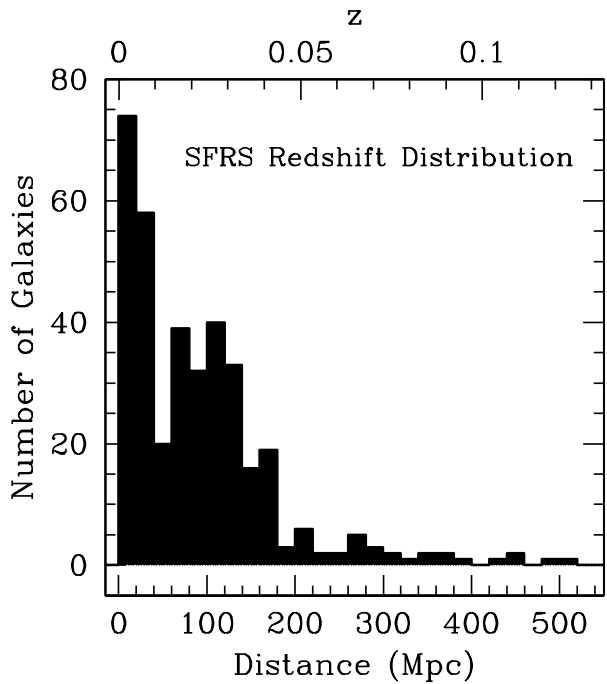}
\caption{
The distance/redshift distribution of the SFRS, in 20\,Mpc bins.
Because of the 60\,$\mu$m \IRAS\ selection, most of the sample galaxies
are nearby ($z<0.05$), but the distribution has a tail of higher-luminosity
galaxies towards higher redshifts.  The plot has been scaled to emphasize
the nearby galaxies; the four most distant objects are not shown (3C\,273,
IRAS\,11267+1558, IRAS\,13218+0552, and OJ\,287, at 660, 740, 850, and
1260\,Mpc, respectively; Table~\ref{tab-basic}).
}
\label{fig:distances}
\end{figure}

The resulting corrected velocities were used to estimate distances for all SFRS 
galaxies, assuming $H_0\,=\,73$\,km s$^{-1}$ Mpc$^{-1}$.
The distances are given in Table~\ref{tab-basic}, and the overall distance 
distribution is plotted in
Figure~\ref{fig:distances}.  The SFRS galaxies
tend to be bright by virtue of their proximity 
(60\% are nearer than 100\,Mpc, and 90\% are closer than 180\,Mpc) 
and are therefore easily accessible from a wide range of telescopes.

\subsection{IRAC Photometry}
\label{ssec:irac}

The IRAC (Fazio \etal 2004) observations of most SFRS sources 
(\ie, all those lacking suitable archival data) were carried out during
Cycle~5 (PID\,50128) using standard observing parameters.
Each of the 273 galaxies targeted by PID\,50128 was observed with at least
$6\times12$\,s full-array exposures using a cycling dither with a 
medium dither scale.  All sources were observed in both IRAC 
fields of view, yielding a total exposure time of at least 72\,s 
in each of the four IRAC bands at 3.6, 4.5, 5.8, and 8.0\,$\mu$m.
This relatively low exposure time
consistently yielded $S/N>1000$ for these very bright galaxies.
All sources not included in PID\,50128 had already been
observed by other programs, and the corresponding data were downloaded
from the archive. 

The archival and PID\,50128 data were reduced together in
as homogeneous a manner as possible.
The basic IRAC data reduction was carried out by team members at the University
of Western Ontario and at the Center for Astrophysics.  In both
instances it was based on the corrected Basic Calibrated Data
(cBCD).  The cBCD frames were object-masked
and median-stacked on a per-AOR (Astronomical Observing Request) 
basis.  The resulting stacked
images were visually inspected and subtracted from individual
cBCDs within each AOR.  This was done to eliminate long-term residual images
arising from prior observations of bright sources by the 3.6, 5.8, and 
8.0\,\micron\ arrays.  The 4.5\,\micron\ detector array did not 
suffer from residual images during \SS's cryogenic mission. 
Subtracting the median stacks also
minimized gradients in the celestial backgrounds around
each source.
After these preliminaries, the data for each galaxy were mosaiced into 
four spatially-registered mosaics using IRACproc (Schuster \etal\ 2006).  
IRACproc augments the capabilities of the standard IRAC reduction software
(MOPEX).  The software was configured to automatically flag and
reject cosmic ray hits based on pipeline-generated masks together
with a sigma-clipping algorithm for spatially coincident pixels.
IRACproc calculates the spatial derivative of each image and adjusts
the clipping algorithm accordingly.  Thus, pixels where the derivative
is low (in the field) are clipped more aggressively than are pixels
where the spatial derivative is high (point sources).  This avoids
downward biasing of point source fluxes in the output mosaics.
The mosaics were resampled to 0\farcs84 per pixel, so each pixel
in the final mosaic subtends half the solid angle of the native IRAC pixels.  

Photometry was carried out with SExtractor (ver. 2.5.0; Bertin \& Arnouts 1996).
Because our interest lies in global photometry and in particular in ensuring
accurate global SEDs, we first registered all IRAC and \GG\ mosaics (for which
\GG\ data were available; Sec.~\ref{ssec:galex}) to a common
spatial scale using SWarp (version 2.17.1) and then applied SExtractor in two-image
mode to photometer the galaxies.  
In two-image mode, SExtractor performs detection and characterization of the
surface brightness distribution in one image and applies that surface brightness
distribution to a second image.  
This was necessary because the infrared and ultraviolet
mosaics in many instances have significant morphological differences;
two-image mode prevents this fact from biasing the global color measurements.
We used the IRAC 3.6\,\micron\ mosaic --- the
best tracer of the stellar light distribution --- as the 
detection image and photometered the \GG\ and IRAC mosaics using the surface
brightness distribution at 3.6\,\micron\ to define the apertures.   In a few
cases where UV-bright features dominate, it was necessary to use the
\GG\ NUV image as the detection image in order to ensure that the SExtractor 
captures all the light in all six \GG\ and IRAC bands.
The IRAC photometry appears in Table~\ref{tab-irac1}.

\subsection{Radio Continuum Flux Densities}

Most of the radio observations come from the NVSS, a snapshot survey of 82\% of the
celestial sphere at 1.4\,GHz (Condon \etal 1998).  The survey's key features are 
that it used VLA
D-array, and the images were cleaned and restored with a 45\arcsec\
beam \citep{Condon1998}.  Typical rms brightness fluctuations in the
survey are 0.45\,mJy\,beam$^{-1}$, and the survey catalog reaches 
the 50\% completeness level at $S = 2.5$\,mJy.  In practice, the SFRS 
sources were retrieved from the online 
catalog\footnote{http://www.cv.nrao.edu/nvss/NVSSlist.shtml, lookup done in
2010 in catalog dated 2004} with a 30\arcsec\ search radius.  Such
queries returned results for 342 of the SFRS galaxies, all with
unique matches.  Figure~6 of \citet{Condon1998} shows a 1\% chance of
finding a spurious match within this distance of an arbitrary
position.  Ten additional matches, two of them double, were found in 
a search radius of 60\arcsec, and two additional matches were found with a
90\arcsec\ search.  The probability of spurious matches at these
distances is 5--10\%, but such large position errors are highly
unlikely for correct matches unless the source itself is very extended.

\citet{Condon2002} gave flux densities based on NVSS data for 192 of
the SFRS galaxies.  In all but 14 cases, the flux densities are
essentially identical to the ones from the automated lookup.  For the
14 discrepant cases and also for all cases where the NVSS position
differed from the adopted galaxy position by more than 10\arcsec, we
examined the NVSS
images\footnote{http://www.cv.nrao.edu/nvss/postage.shtml} together
with the \IRAS\ and IRAC infrared images.  Most of the discrepancies
were caused by blends of two radio sources, one of them being the SFRS 
galaxy and the second a nearby galaxy or QSO.  This cause (blended sources) 
was found to explain discrepancies both in total measured flux and for 
discrepancies in the reported coordinates. 
In about half of the cases, the blended
source was identifiable on the IRAC image or in NED.  Most of the
time, the \IRAS\ (\ie, SFRS) source was attributable to a single
galaxy, but in some cases two galaxies are likely to contribute.  
The radio blends were deconvolved using {\sc imfit} in the Common
Astronomy Software Applications (CASA)
package\footnote{http://casa.nrao.edu/index.shtml}, in each case
with the minimal set of free parameters that led to a satisfactory
fit\footnote{When small-diameter objects could be identified, their
positions were held fixed, and objects were forced to be point
sources whenever possible.  Extended objects had position angles held
fixed when those could be determined from visible or IR
data.}.  Another cause of discrepancy was extended radio sources.  The
NVSS catalog fit Gaussians to all sources, but in some cases this is
not an accurate description of the galaxy.  In these cases, the radio
flux was measured in a circular, rectangular, or polygonal beam as
appropriate for the particular galaxy. In general, our results agree
with those of \citet{Condon2002} but differ slightly because we took
into account information from the \IRAS\ images in determining how to
fit the radio sources.

A final 16 SFRS sources either were not detected in the NVSS or had
very low signal to noise.  These were observed with the VLA in D
configuration in 2008 July (program AA~319).\footnote{The AA~319
observations also included two galaxies that were in an early version
of the SFRS sample but were later deleted.}  The observations used
the same observing frequencies and bandwidth as those of the NVSS but
were $\ge$7~minutes long as opposed to 30~s for the NVSS in the
relevant declination bands.  Data reduction was with CASA and
followed the NVSS procedure in using superuniform weighting with
{\tt npix}$=5$. The reductions differed in using 7\farcs5 pixels
and a restoring beam set by the actual baselines and weightings of
each observation; typical beams were 30\arcsec$\times$40\arcsec. The
two frequencies observed were imaged and cleaned separately and the
resulting images averaged; the separate images were also inspected
and source flux densities measured on each of them as well as on the
combined image. This was especially important for three galaxies with
bright radio sources (M87, 3C~273) in the outer part of the VLA
primary beam. Flux densities were measured via Gaussian fitting (CASA
{\sc imfit}), by adding up pixels in rectangular areas, or where
needed by deconvolution as discussed above.  In many
cases, the limiting noise source is sidelobes from strong,
imperfectly cleaned sources in the field.  The tabulated
uncertainties are our best estimate taking these into account,
especially by comparison of the images at the two frequencies and of
different methods of measuring flux density.

All the radio flux densities are given in Table~\ref{tab-radio}.  Uncertainties
are higher for sources marked as ``extended'' because the peak
surface brightness is the best-measured parameter, and uncertainty in
the source size contributes to the uncertainty in flux density.
Unrecognized blends may give spuriously high flux densities, but this
is likely to be the case for only a very few objects.

\subsection{MIPS 24\,$\mu$m Photometry}

Many SFRS galaxies are not well-detected by \IRAS\ at 12 and 25\,$\mu$m,
leaving an obvious gap in the suite of useful SFR estimators (\eg, Calzetti \etal 2010). 
To fill this gap we photometered all available archival 
\SS/MIPS observations (Rieke \etal 2004) and were awarded time to observe 
the remainder of the sample via our own observing program (PID\,50132, PI Fazio).  

Our MIPS 24\,$\mu$m campaign was active from the start of 2008 November until the
exhaustion of \SS's cryogen in 2009 May.  A total of 178 SFRS galaxies were observed,
all of them in Phot mode.  Most used a small field size, 10\,s exposures per position, 
and 2 cycles.  For six relatively large galaxies, a large field size was used to 
ensure a sufficiently large source-free background was present in
the final mosaics to permit an accurate background subtraction.  In one instance
(NGC\,3338), a raster map had to be used to cover the full spatial extent of the source
plus the nearby field.

All PID\,50132 data were reduced using standard techniques.  We used object-masked 
median stacks of all exposures of each target to eliminate array artifacts from
the enhanced basic calibrated data (eBCD) before mosaicing.  All mosaics were pixellated to 2\farcs5.

The archival observations were a heterogenous dataset employing a variety of
exposure times and observing strategies.  Our analysis of these 101 observations 
therefore began with the post-BCD data products, which retain some array-based
artifacts but were nonetheless suitable for deriving global photometry.
Two galaxies (NGC\,4314 and 4418) exhibited saturation at their cores in the
archival data, but apart from this we retrieved mosaics covering the full
extent of a total of 101 of our sample galaxies.  Together with the 178 objects
from PID\,50132, a total of roughly 3/4 of the SFRS galaxies yielded useful mosaics.

We photometered all sources identically using SExtractor, accounting for differences
in pixellation by an appropriate choice of convolution kernel.  Typically the sources
were detected with $S/N$ ratios of hundreds or even thousands in mosaics that were
very far from being confused with unrelated background or foreground sources. 
Using the effective radii measured by SExtractor ({\tt KRON\_RADIUS}), we 
applied appropriate aperture corrections to the total
fluxes following Table~4.13 of the MIPS Instrument Handbook.  
No special effort was made to identify and exclude foreground stars, but because 
of the typically very high galactic latitude of the sources this should not 
significantly bias the photometry.

The MIPS photometry was verified in two ways.  First, all the SExtractor-generated 
background and object checkimages were inspected to make sure that the
backgrounds were smooth on scales larger than the galaxies themselves and that
SExtractor had identified all the pixels associated with a particular galaxy.
Second, the results were compared to those reported for the 15 SFRS galaxies 
in common with the SINGS and LVLS samples (Table~\ref{tab:sings}; 
Dale \etal 2007; Dale \etal 2009).
All agree within $1\sigma$ except for NGC\,5474 and NGC\,4395. 
The measurement for NGC\,5474 (0.14\,Jy) is 1.7$\sigma$ lower than the 
SINGS measurement (0.18\,Jy).  Among 15 independent measurements, a 
single discrepancy at the 1-2$\sigma$ level is acceptable agreement.  
NGC\,4395 is arguably a different case because the photometry is discrepant 
at the 3$\sigma$ level.  
This galaxy is unusually extended with a particularly uneven surface brightness
profile.  The failure of the curve-of-growth to converge on a single level strongly
suggested the MIPS photometry for NGC\,4395 was biased low.  
For this single object the Dale \etal (2009) 
measurement was therefore adopted in preference to our own measurement.

The final aperture-corrected global MIPS 24\,$\mu$m photometry is 
given in Table~\ref{tab:fir}.
The corresponding aperture corrections are given in Table~\ref{tab-mips}.
Because the detections have such high $S/N$ ratios, the random (measurement) uncertainty
is dominated by the systematic uncertainty in the absolute calibration of MIPS.
According to the MIPS Instrument Handbook this uncertainty is between 4 and 8 percent;
we therefore adopt 8\% uncertainty as the most conservative estimate of the true
error in all the MIPS photometry.
For reasons having to do purely with scheduling, the galaxies lacking MIPS 24\,$\mu$m
photometry tend to be those at lower Right Ascensions; eventually this lack will be
corrected using photometry from the WISE mission.

\subsection{Far-Infrared Photometry from Planck}

{\sl Planck} (Planck Collaboration 2011a) is a space-based mission now carrying 
out an all-sky survey in six bands from 25--1000\,GHz with a spatial resolution 
that progresses from $\sim$30\arcmin\ to 5\arcmin\ depending on the band.  
Although its main mission is to measure spatial anisotropies in the Cosmic Microwave 
Background, in the course of its repeated surveys of the sky it has detected thousands 
of foreground sources.  The 1700+ detections resulting from the first pass through the sky 
are tabulated in the {\sl Planck} Early Release 
Compact Source Catalog (Planck Collaboration 2011b).  Because the SFRS galaxies are 
bright, a significant number are detected by {\sl Planck}: 176, 78, and 28 are detected 
by {\sl Planck's} High-Frequency Instrument (Lamarre \etal 2010)
at 350, 550, and 850\,$\mu$m, respectively.  This far-infrared photometry provides new and
valuable constraints on the cold dust content of the detected SFRS galaxies, \eg, Planck Collaboration (2011c)
found evidence for cold ($T<20\,K$) dust in their analysis of combined {\sl IRAS}+{\sl Planck} SEDs.
The relevant photometry is presented in Table~\ref{tab:fir}.  No doubt the SFRS detection
fraction will increase as {\sl Planck} accumulates more complete passes over the sky and 
reaches correspondingly fainter flux limits.

\subsection{Visible and Near-Infrared Observations}

The visible photometry was taken from the data release 7 of the Sloan digital sky survey
\citep[SDSS;][]{abazajian09}. The SDSS consists of an imaging
survey of $\pi$ steradians, mainly in the northern sky, in five passbands
$u, g, r, i$, and $z$. The imaging was done in drift-scan mode, and the data were
processed using the photometric pipeline \textsc{photo} \citep{lupton01}
specially written for SDSS.  
All SFRS galaxies have photometry available in the five SDSS bands, 
although in six cases the cataloged values greatly understate the true values 
because they pertain only to the galaxy nuclei instead of the entire galaxy.
We used the Petrosian magnitudes as the best measure of the total flux 
in the five SDSS bands \citep{blanton01}.  The Petrosian magnitudes were
calculated using the aperture set by the `Petrosian radius' in the $r$-band, thus
providing consistent measurements.  SDSS Petrosian magnitudes should
recover all the flux for an exponential galaxy profile independent of the axis ratio
\citep{blanton01} and about 80\% of the flux for a de Vaucouleurs profile.  

SDSS also acquired nuclear spectra for 210 SFRS galaxies.
The spectra were obtained using two
fiber-fed double spectrographs covering a wavelength range of
3800--9200 \AA. The resolution $\lambda/\Delta\lambda$ varies between
1850 and 2200.  The SDSS fibers have a diameter of 3\arcsec.

Near-infrared photometry was taken from the 2MASS extended source catalog.
Although isophotal magnitudes had been used for the initial SFRS sample selection, 
total magnitudes were extracted from the 2MASS database for greater consistency 
with the global photometry measured in the other bands.  
2MASS extrapolated total magnitudes are typically $\sim0.3$\,mag brighter than 
isophotal magnitudes.  Typical measurement uncertainties are given as 0.03, 0.04, 
and 0.05\,mag at $J$, $H$, and $K_s$, respectively.  Sec.~\ref{ssec:pairitel}
discusses the {\sl systematic} biases in the 2MASS photometry.

The SDSS and 2MASS photometry is presented in Table~\ref{tab-phot1}.

\subsection{\GG\ Photometry in the Near- and Far-Ultraviolet Bands}
\label{ssec:galex}

\GG\ (Martin \etal 2005) is an astronomical satellite with sensitive
wide-field ultraviolet imaging capability in two bands, the FUV 
(1350--1750\AA) and NUV (1750--2800\AA).  The \GG\ archive (release GR6) 
contains scientifically usable imaging (covering the full extent of our 
sources and therefore capable of yielding global flux measurements) 
in at least one  of the \GG\ bands for 332/369 SFRS galaxies, or 
90\% of the sample.  The fraction observed 
by \GG\ in the NUV will increase as the survey portion of the mission continues,
although FUV imaging is no longer possible.

Almost three-quarters of these imaging data in both wavebands were taken as 
part of \GG's primary surveys\footnote{{\tt http://galexgi.gsfc.nasa.gov/docs/galex/ Documents/ERO\_data\_description\_2.htm}} -- 
the all sky survey (AIS) with an effective exposure time of 
$\sim\!0.1$\,ks and the relatively deep nearby galaxy survey (NGS) with 
an effective exposure time of $\sim\!1.5$\,ks.  With a resolution of 
$4^{\prime\prime}\!-\!6^{\prime\prime}$, \GG\ images have lower spatial 
resolution than \SS/IRAC.  However, since almost a third of the SFRS 
galaxies lie $<\!40$\,Mpc away, together with a 1.25 degree \GG\  
field of view, the moderate \GG\ spatial resolution nonetheless allows for 
robust measures of spatially-integrated fluxes and color which are comparable 
to data at other wavelengths.  The  \GG\ observations and an analysis of 
the UV properties of the SFRS galaxies are described by S. Mahajan \etal 
(2012, in preparation; hereafter Paper~2).

\begin{figure*}
\includegraphics[width=180mm, height=85mm]{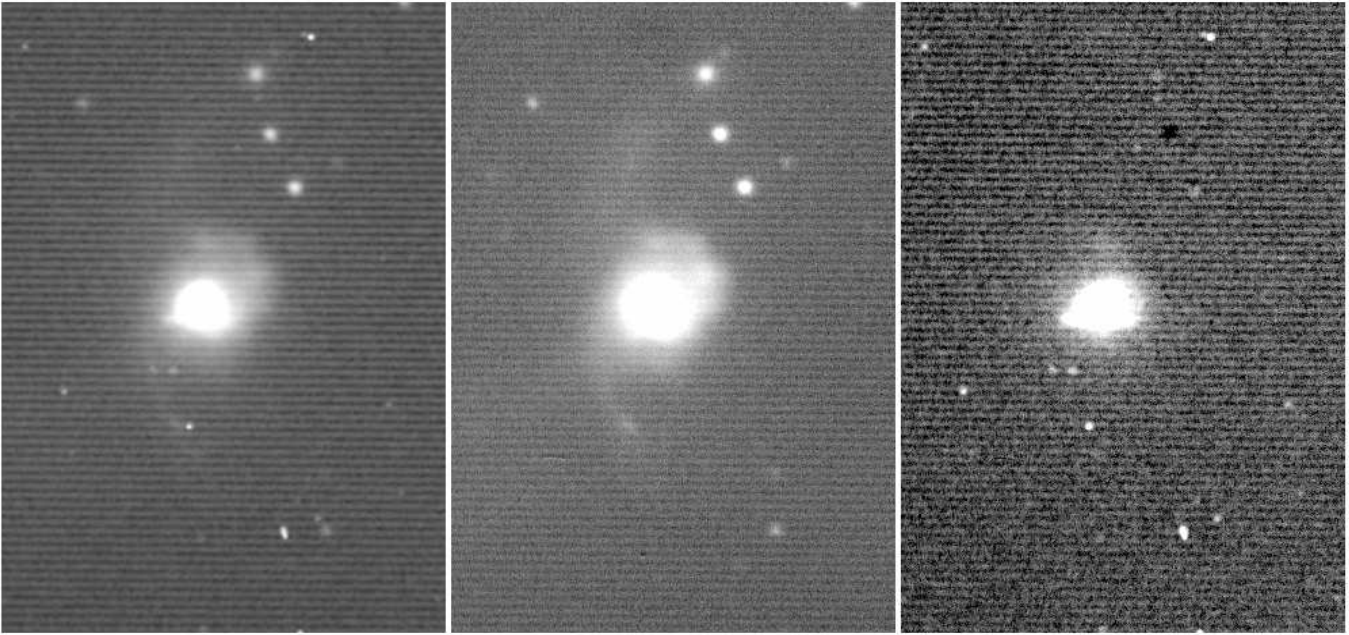}
\caption{
A typical example of the narrowband H$\alpha$ imaging being collected for the
SFRS sample; the images shown here are for UGC\,8058.
{\sl Left panel:} the narrowband H$\alpha$ emission-line
image obtained at the NOAC.  {\sl Center panel:} the same field, seen in the
continuum at $R$.
{\sl Right panel:} a difference image formed by subtracting the continuum image
from the narrowband image, revealing the regions in which significant
star formation is traced by the H$\alpha$ emission.
}
\label{fig:halpha}
\end{figure*}

\subsection{H$\alpha$ Imaging from the National Astronomical Observatory of China}

In the spring of both 2008 and 2009, \halpha\ imaging was acquired for SFRS sample 
galaxies from the NAOC 2.16\,m telescope in Xinlong, Hebei Province, China.
The Beijing Faint Object Spectrograph and Camera (BFOSC) was used to
obtain on-source integrations lasting between 1800 and 3600\,s for each
galaxy observed. 
The choice of exposure time was made based on prevailing conditions;
longer exposures were used during bright-sky time or for relatively faint targets.   
The 10\arcmin$\times$10\farcm5 BFOSC field of view illuminates a single CCD with
2048$\times$2080 pixels.  Each pixel subtends roughly 0\farcs3, well below the typical
seeing during our campaign (2\farcs0 -- 2\farcs5).  This spatial resolution is reasonably 
well-matched to the FWHM intrinsic to  IRAC (1\farcs66 -- 1\farcs98).  The BFOSC
is equipped with a suite of 11 narrow \halpha\ filters centered at wavelengths ranging 
from 6563\,\AA\ to 7060\,\AA\ at 50\,\AA\ intervals.  The filter bandpasses are 70\,\AA\ wide.
In the two observing campaigns carried out to date, a total of 105 SFRS galaxies were 
observed through one of these narrow-band \halpha\ filters, the specific filter being 
chosen to cover the \halpha+[NII] complex at the target's redshift.  Observations were
also made of each galaxy through the standard Johnson $R$ band filter but
with shorter exposure times (400 -- 600\,s) in order to measure the continuum flux.

The BFOSC data were reduced in the usual way.  The lowest 32 rows of pixels were used
for overscan correction.  Standard IRAF tasks were used to subtract bias frames, 
construct and apply flatfields, coadd frames while simultaneously correcting for cosmic 
rays and known bad pixels using outlier rejection, and finally to photometer each of the 
105 observed sources in both \halpha\ and $R$.  This work was carried out at the Chinese 
Academy of Sciences in Beijing.  The quality of the final \halpha\ mosaics is
indicated by Fig.~\ref{fig:halpha}, which shows the outcome for UGC\,8058, a typical case.

\begin{figure*}
\includegraphics[width=175mm]{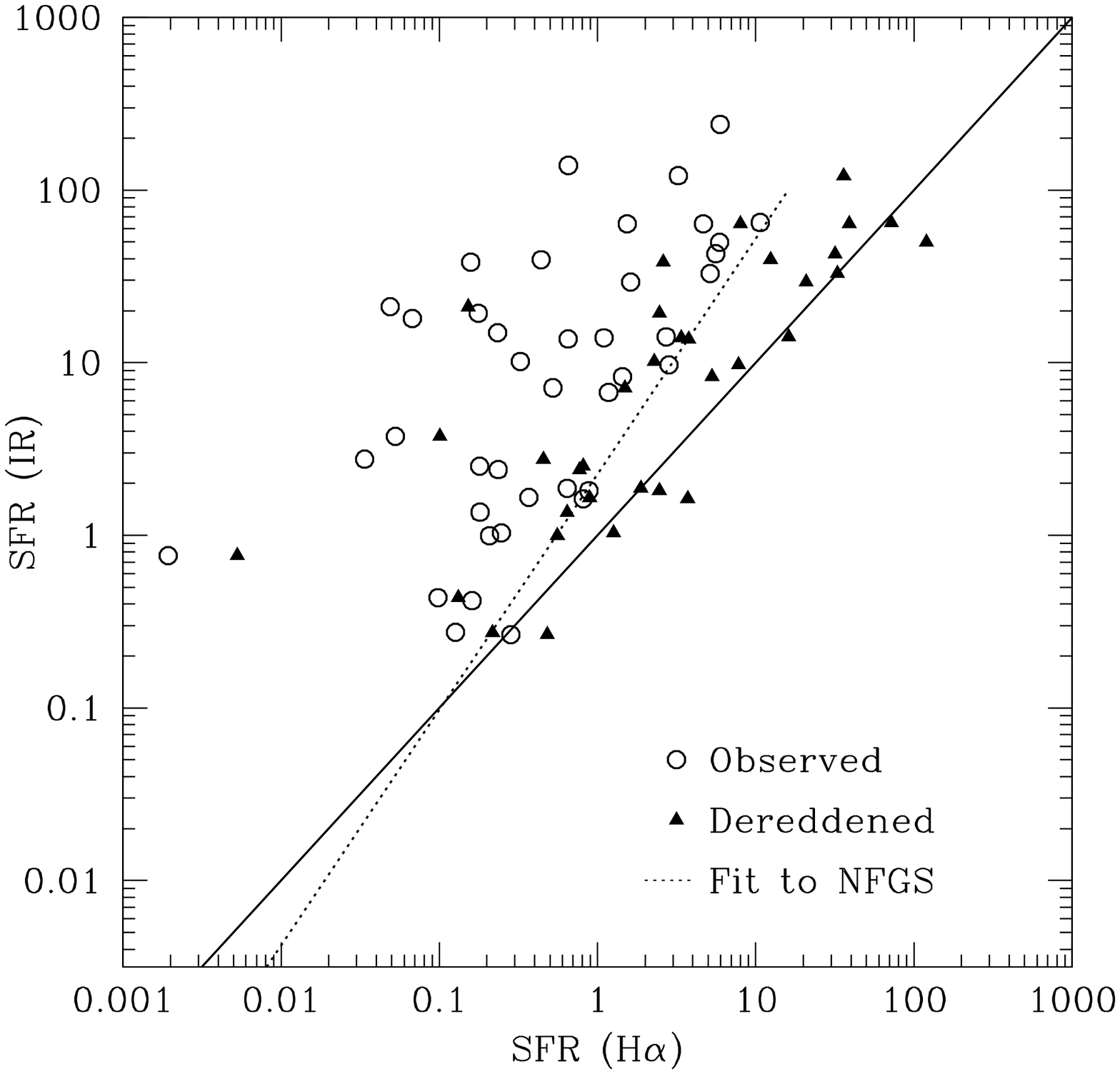}
\caption{
A comparison of far-infrared and \halpha\ SFRs for an unrepresentative subset
of the SFRS, following Fig.~1 of Kewley \etal\ (2002).
Circles denote galaxies when SFR(\halpha) is not correct for extinction due
to dust.
Solid triangles show the revised (larger) SFR(\halpha) estimates that result
when the extinction estimates from Paper~2 are applied.
The solid line indicates
where the points would fall if the Kennicutt (1998) relations were in agreement.
The dashed line indicates the Kewley \etal\ (2002) fit to the corresponding
data for the NFGS, not extinction-corrected.
}
\label{fig:kewley}
\end{figure*}

Figure~\ref{fig:kewley} shows the relationship between the SFRs derived independently 
in \halpha\ and the far-infrared for the 40 SFRS galaxies fully processed so far.
The SFR estimates were made following the methods applied by Kewley \etal\ (2002) 
to 81 `pure' starburst galaxies in the Nearby Field Galaxy Survey 
(hereafter NFGS).  Specifically, we estimated the \halpha-based SFR as
${\rm SFR}(\halpha) = 7.9\times10^{-42} L(\halpha)$ 
and the far-infrared SFR as ${\rm SFR(FIR)} = 4.5\times10^{-44} L(TIR)$ (Kennicutt 1998), 
where we have estimated the total far-infrared luminosity $L(TIR)$ as described 
in Sec.~\ref{sec:agn}.

Kewley \etal\ (2002) demonstrated that for 81 NFGS galaxies selected 
at visible wavelengths to cover a range of luminosities, SFR(\halpha) is discrepant with 
SFR(IR) in the sense that SFR(\halpha) 
tends to be significantly lower than SFR(IR).  The discrepancy was measured to be
a factor of nearly three (2.7$\pm0.3$) and was shown to increase 
with increasing SFR.  Both of those findings apply also to the 
SFRS galaxies shown in Fig.~\ref{fig:kewley}, but more so: as shown subsequently, the 
infrared-selected SFRS galaxies require a significantly larger mean reddening 
correction than did the NFGS galaxies to bring the \halpha\ and far-infrared SFR 
indicators into agreement.

Most of the galaxies shown in Fig.~\ref{fig:kewley} have \GG\ photometry available.
For all such objects an extinction correction $A_{H\alpha}$ appropriate for \halpha\ 
was estimated based on the NUV photometry in Paper 2, which implemented 
the empirical IRX-based prescription following Eq.\,1 of Buat \etal (2005).  
That estimate for $A_{NUV}$ was then converted into an \halpha\ 
extinction estimate $A_{H\alpha}$ assuming a Calzetti (2001) dust law:
$ A_{H\alpha} = 0.79 (A_V + A_{MW}) = 0.79 (0.77\times A_{NUV}+ A_{MW})$
where $A_{MW}$ is the foreground extinction due to the Milky Way.
Following this prescription the mean A$_{H\alpha}$ was found to be 1.7 mag, 
corresponding to attenuation by a factor of 4.7.  This likely reflects the fact 
that our sample is (primarily) infrared-selected, unlike the NFGS.  
A similar effect is seen for ultraviolet selection (Buat \etal\ 2005).
Intriguingly, four galaxies (out of 34 with extinction estimates) lie 
more than an order of magnitude from the line of equality even when 
extinction-corrected.  This suggests that for a minority of 
infrared-selected galaxies, either the intrinsic \halpha\ 
emission does not accurately reflect the `true' SFR. 
Alternately, the standard extinction corrections may not work well for such 
objects, as has been found for galaxies with large SFR(IR)/SFR(UV)
ratios (Wuyts \etal 2011).
The reason will remain unclear pending acquisition and analysis 
of \halpha\ emission line intensities for the full SFRS.  This will be possible 
as soon as the NAOC \halpha\ imaging campaign is completed.  In the summer of 2010 
the BFOSC was upgraded with a new CCD 
having lower read noise.  A subset of our team (led by H.~Wu) has been awarded time 
in the 2011 observing season to complete the imaging campaign for the 
roughly two-thirds of the SFRS for which \halpha\ imaging has not yet been obtained. 

We have also begun a longslit spectroscopic campaign (Sec.~\ref{sec:fast}) 
to obtain dust reddening estimates for all SFRS galaxies via the Balmer 
decrement and thereby address the possibility that $A_{NUV}$ may not be the 
most reliable extinction indicator at visible wavelengths (Bell 2003).  
A detailed analysis of the global \halpha\ emission line flux 
measurements for the full sample will be presented in a forthcoming paper 
(Y.-N.~Zhu \etal 2012, in preparation).

\section{Discussion}
\label{sec:discussion}

\subsection{SFRS Compared to Other Nearby Galaxy Samples}
\label{sec:comparison}

A number of recent survey programs have documented important aspects of star 
formation phenomenology.  In addition to the studies mentioned above (SINGS, LVLS, 
and MESS), there are the NFGS (Kewley \etal 2002) and the Herschel Reference
Survey (Boselli \etal 2010).  The SFRS differs from these other efforts in
significant ways.

SINGS is fundamentally different than SFRS because SINGS is comprised of a 
relatively small number (75) of very nearby, extended objects.  
SINGS therefore allows spatially-resolved 
measurements in numerous wavebands (Dale \etal 2007), 
but the SINGS galaxies sample primarily the low-luminosity end of the far-infrared
luminosity function.  This means SINGS isn't representative of 
star formation generally. 
Only five SINGS galaxies are present in the SFRS (Table~\ref{tab:sings}).

\tablenum{6}
\begin{deluxetable}{ccc}
\tablecolumns{3}
\tabletypesize{\scriptsize}
\tablecaption{Galaxies in Common with Other Surveys\label{tab:sings}}
\tablewidth{0pt}
\tablehead{
\multicolumn{3}{c}{Spitzer Infrared Nearby Galaxy Survey }\\
\multicolumn{3}{c}{(SINGS) }\\
}
\startdata
NGC\,3049 & NGC\,3265 & NGC\,5474 \\
NGC\,3190 & NGC\,3773 &           \\
\cutinhead{Local Volume Legacy Survey (LVLS)}
NGC\,2500 & NGC\,4020 & NGC\,5474 \\
NGC\,2537 & NGC\,4244 & NGC\,5585 \\
NGC\,2552 & NGC\,4395 &           \\
NGC\,3274 & NGC\,5204 & \\
\cutinhead{Herschel Reference Survey (HR)}
NGC\,3245 & NGC\,4237 & NGC\,4548 \\
NGC\,3338 & NGC\,4294 & NGC\,4592 \\
NGC\,3370 & NGC\,4396 & NGC\,4607 \\
NGC\,3430 & NGC\,4412 & NGC\,4630 \\
NGC\,3659 & NGC\,4420 & NGC\,4689 \\
NGC\,3666 & NGC\,4424 & NGC\,4688 \\
NGC\,3686 & NGC\,4435 & NGC\,4701 \\
NGC\,3729 & NGC\,4438 & NGC\,4747 \\
NGC\,4116 & NGC\,4470 & UGC\,8041 \\
NGC\,4178 & NGC\,4491 & NGC\,5014 \\
NGC\,4207 & NGC\,4519 & NGC\,5303 \\
\enddata
\end{deluxetable}

The distinction between the LVLS and SFRS is less obvious
because of the two-tiered optical+volume-limited nature of the selection 
used for the LVLS.  The fundamental difference is that 
the LVLS is mostly comprised of low-luminosity dwarf galaxies, well below
the break in the far-infrared luminosity function at $L_{FIR}=10^{10.25}$\,\Lsun.  
Even so, the overlap between the LVLS and the SFRS samples
is surprisingly small --- only 10 objects (Table~\ref{tab:sings}).

The MESS sample relies on a 
SDSS \halpha\ emission-line strength selection criterion to define a sample of 138 
IR-luminous galaxies that are very rapidly forming stars.  MESS galaxy SFRs 
range from 11 to 61\,\Msun\,yr$^{-1}$.  Because it uses a single
selection criterion, MESS does
not control for ISM temperature or stellar mass and so cannot 
be representative of even high-intensity star formation generally. 
Furthermore, \halpha\ fluxes systematically understate total
SFRs compared to the far-infrared unless relatively 
large and uncertain extinction corrections are made, and MESS selects against
star-forming galaxiess with large extinction.  By design, MESS samples only
galaxies with large SFRs.

GOALS (Armus \etal 2009) is more akin to the SFRS because it relies on a restricted 
60\,$\mu$m brightness selection to construct the sample.  All of the more than 200 GOALS galaxies
are likewise bright, nearby objects for which an impressive range of multiband
data have been made available, particularly HST and CXO imaging.  But by design, GOALS 
selects for Luminous InfraRed Galaxies (LIRGs, galaxies with $L>10^{11}$\,\Lsun).  
The AGN fraction is significantly higher for high-luminosity galaxies.
Indeed, it is a feature of the GOALS sample that it contains the full range of 
optical spectral types including many dominant AGNs as well as major-merger systems.
While SFRS galaxies are by no means entirely free of AGN contributions 
(Sec.~\ref{sec:agn}), they are a relatively minor contributor, 
and the fact that the SFRS 
spans the full range of both far-infrared color and luminosity simultaneously makes the 
SFRS less biased in characterizing star formation than GOALS.

{\sl The \HH\ Reference Survey} (Boselli \etal 2010)
contains nearly as many galaxies as the SFRS (323 vs.~369) but does 
not capture the full range of star forming behavior because it is explicitly
oriented toward high-density environments.
Consequently, it contains a much larger fraction of elliptical 
galaxies (65/323) residing in Virgo and other major clusters.  The \IRAS\ 60\,$\mu$m
detection fraction is about 3/4, primarily because of a low detection
fraction for HRS ellipticals.  
HRS contains 33 galaxies in common with the SFRS
(Table~\ref{tab:sings}), an overlap of about 10\%.  
Most important, HRS approximates two morphology-based volume-limited
surveys, together spanning `only' three decades in 60\,micron\
luminosity.  HRS has or soon will have a comprehensive
suite of \HH\ far-IR photometry available, but its selection makes it
optimized for the study of how environment affects dust 
content rather than as an unbiased study of star formation.

\subsection{AGN Content}
\label{sec:agn}

The SFRS does not discriminate for AGN activity.  For this reason it should be an 
excellent means of estimating AGN prevalence generally as well as the relative importance 
of AGNs for the overall energy budget of any far-IR-selected sample, at 
least in the local Universe.  Our ongoing campaign (Sec.~\ref{sec:fast}) 
to acquire longslit spectra for all SFRS galaxies with the 
FAst Spectrograph for the Tillinghast Telescope (FAST; Fabricant \etal 1998) 
will eventually permit AGN detection 
to very faint levels via resolved emission line intensity ratios (\eg, Kewley \etal 
2006).  But even though the faint AGNs will be out of reach or a time, it is 
nonetheless possible to identify the relatively luminous AGNs immediately via their 
global infrared colors, and to exploit the spectroscopy that does exist to arrive
at a preliminary estimate of AGN prevalence.

\begin{figure}
\includegraphics[width=90mm]{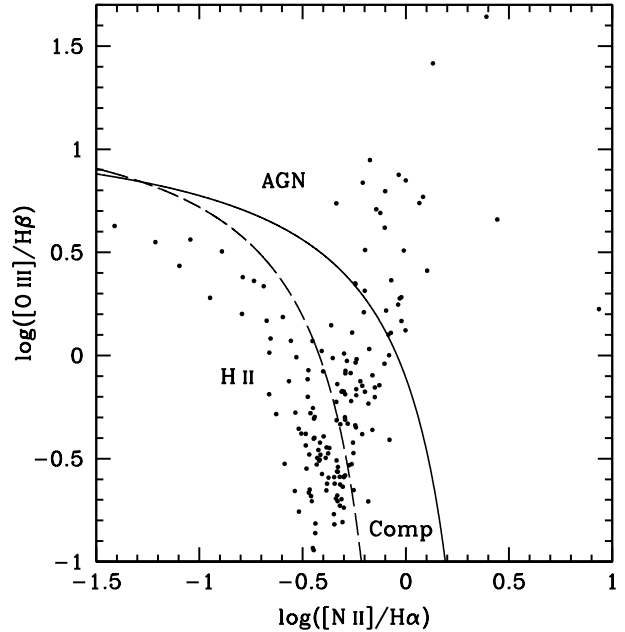}
\caption{
Line ratios from SDSS fiber spectra for 165 SFRS galaxies.  Galaxies
above the solid line are considered AGNs \citep{Kewley2001}, and
those below  the dashed line are considered purely star forming
\citep{Kauffmann2003}, with emission-line intensity ratios closely
matching those seen in HII regions.  Those between the two lines are
composite or transitional.}
\label{f:bpt}
\end{figure}

The most sensitive method for identifying AGNs is probably the 
BPT diagram \citep{Baldwin1981,Kewley2001}, an emission-line intensity ratio
diagnostic that identifies AGNs via species that trace
high-excitation conditions.  Because it relies on visible lines, 
the BPT method will not find highly-obscured AGNs, but it can
identify so-called `transition' sources in which the contributions of
star-forming activity and a central AGN are comparable.
Figure~\ref{f:bpt} shows this method applied to a subsample of
SFRS galaxies, \ie, all 165 objects for which the set of four
necessary emission line intensities have been measured by SDSS.
Of these 165 sources, 30 are identified as AGNs according to the 
\citeauthor{Kewley2001} line-ratio criterion.  Another 45 objects
were observed spectroscopically by SDSS but do not exhibit the emission
lines linked with AGNs.  These 45 galaxies are unlikely to host strong
AGNs.  Thus the AGN fraction identified by the BPT diagram is roughly 1/7.

\begin{figure}
\includegraphics[width=90mm]{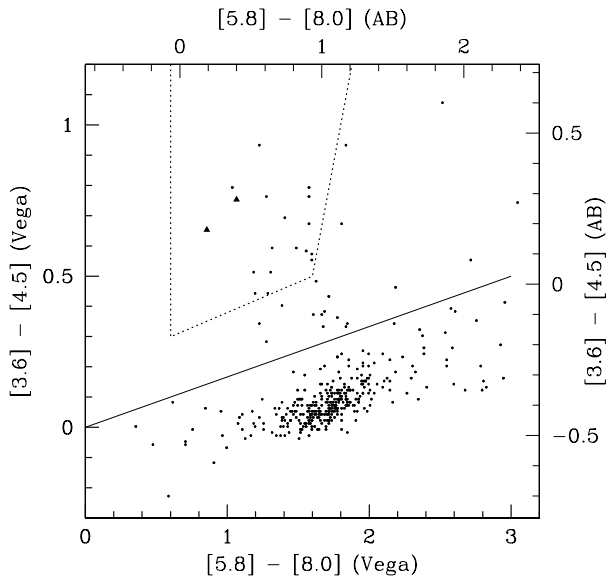}
\caption{IRAC colors of SFRS galaxies.  Colors are shown in both
Vega and AB magnitudes on opposite axes.  The solid triangles denote
3C\,273 and OJ\,287.  The dotted line encompasses the empirically-determined region
within which galaxy colors indicate the presence of an AGN \citep{Stern2005}.
The solid line indicates the division between normal galaxies and outliers,
which probably represent a mix of AGNs and strongly star-forming systems.
}
\label{fig:stern}
\end{figure}

Figure~\ref{fig:stern} shows the IRAC color-color plot first used by 
Stern \etal (2005) to empirically discriminate galaxies with infrared SEDs 
dominated by active nuclei from those dominated by star formation.  
It identifies 19 of the 369 SFRS galaxies ($\sim$5\%) as AGNs according to 
the empirical Stern et al.~criteria. 
For these sources, emission from an active nucleus is a 
significant or even dominant contributor to the mid-IR SED.  
Moreover, some of the 18 galaxies outside the Stern et al. ``wedge'' 
but having redder 
$[3.6]-[4.5]$ colors than the main group probably also contain AGNs. 
The most extreme such outliers are UGC\,9560 (=II\,Zw\,70) and
UGC\,5613, which have very different luminosities but are both strongly
star-forming \citep{Oconnell1978, Poggianti2000}.  
An outlier with much smaller $[5.8]-[8.0]$ color is IC\,486 (=UGC\,4155), 
which is a Sey\,1 \citep{Bonatto1997}.
The outliers are thus a mixed group.

\begin{figure}
\includegraphics[width=88mm]{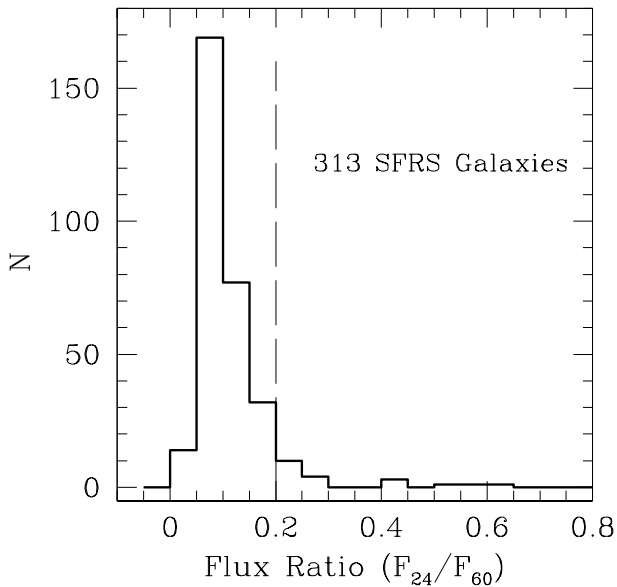}
\caption{
The distribution of the ratio $F_{24}/F_{60}$
for 313 SFRS galaxies for which MIPS 24\,$\mu$m
fluxes (or their \IRAS\ 25\,$\mu$m equivalents) are available.
The vertical line marks the value above which Sanders \etal (1988) suggest
that a relatively large mid-infrared flux indicates that an AGN contributes
to the SED. Out of 313 SFRS galaxies 22 meet this criterion.
They are listed in Table~\ref{tab:agn_frac}.
}
\label{fig:fluxratio}
\end{figure}

A third method to identify AGNs is based on the 25~\micron\ (or 24~\micron) 
flux density \citep{Sanders1988}, specifically $F_{25}/F_{60} > 0.2$.  
Despite their large 60\,$\mu$m fluxes, many SFRS galaxies are not reliably
detected by \IRAS\ in the 25\,$\mu$m band.  We therefore used MIPS 24\,$\mu$m flux
measurements, where available, to form the flux density ratio $F_{24}/F_{60}$,
because MIPS 24\,$\mu$m fluxes track \IRAS\ 25\,$\mu$m
fluxes very closely on average (Dale \etal 2009).  
The distribution of the 313 SFRS galaxies with such data
is shown in Figure~\ref{fig:fluxratio}.  
The distribution is qualitatively similar to that of the Revised
Bright Galaxy Survey (Sanders \etal 2003).
Only a small minority of the sources (22/313, or about 7\%) lie at $F_{24}/F_{60} > 0.2$, 
which Sanders \etal (1988) identify as potentially arising from a significant
contribution coming from an active nucleus.  
Most of the galaxies identified as AGNs in this way are also identified by one of the
other two methods (Table~\ref{tab:agn_frac}). 

\begin{figure}
\includegraphics[width=90mm]{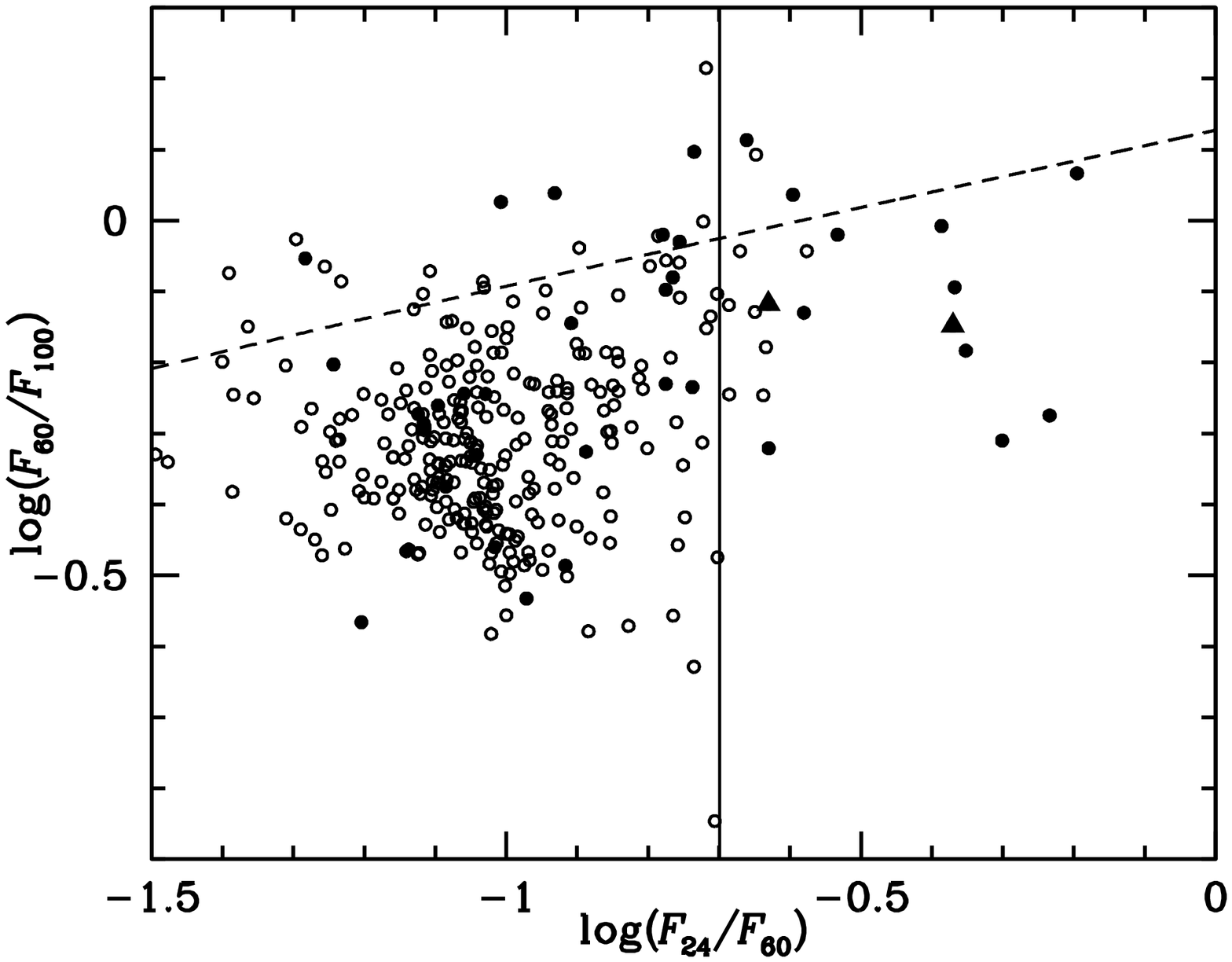}
\caption{
Far-infrared color-color diagram for 313 SFRS galaxies. The
24~\micron\ flux densities come from MIPS if available or
{\sl IRAS} (25~\micron) otherwise, and the 60 and 100~\micron\ flux
densities come from {\sl IRAS}\null.  Galaxies to the right of the
vertical line meet the \citet{Sanders1988} criterion for an active
nucleus that affects the far-infrared colors.  The dashed line indicates the
colors of isothermal dust with emissivity $\epsilon \propto \nu^{-1}$.
Filled symbols indicate galaxies identified as AGN on the basis of nuclear SDSS
spectra (Fig.~\ref{f:bpt}) or the Stern diagram (Fig.~\ref{fig:stern}).
The solid triangles correspond to 3C\,273 and OJ\,287.
The 56 SFRS galaxies having only upper limits at 25~\micron\ are not plotted,
but only 27 of them have 25~\micron\ upper limits large enough
to potentially allow them into the AGN territory.
}
\label{fig:agn_iras}
\end{figure}

\begin{figure}
\includegraphics[width=90mm]{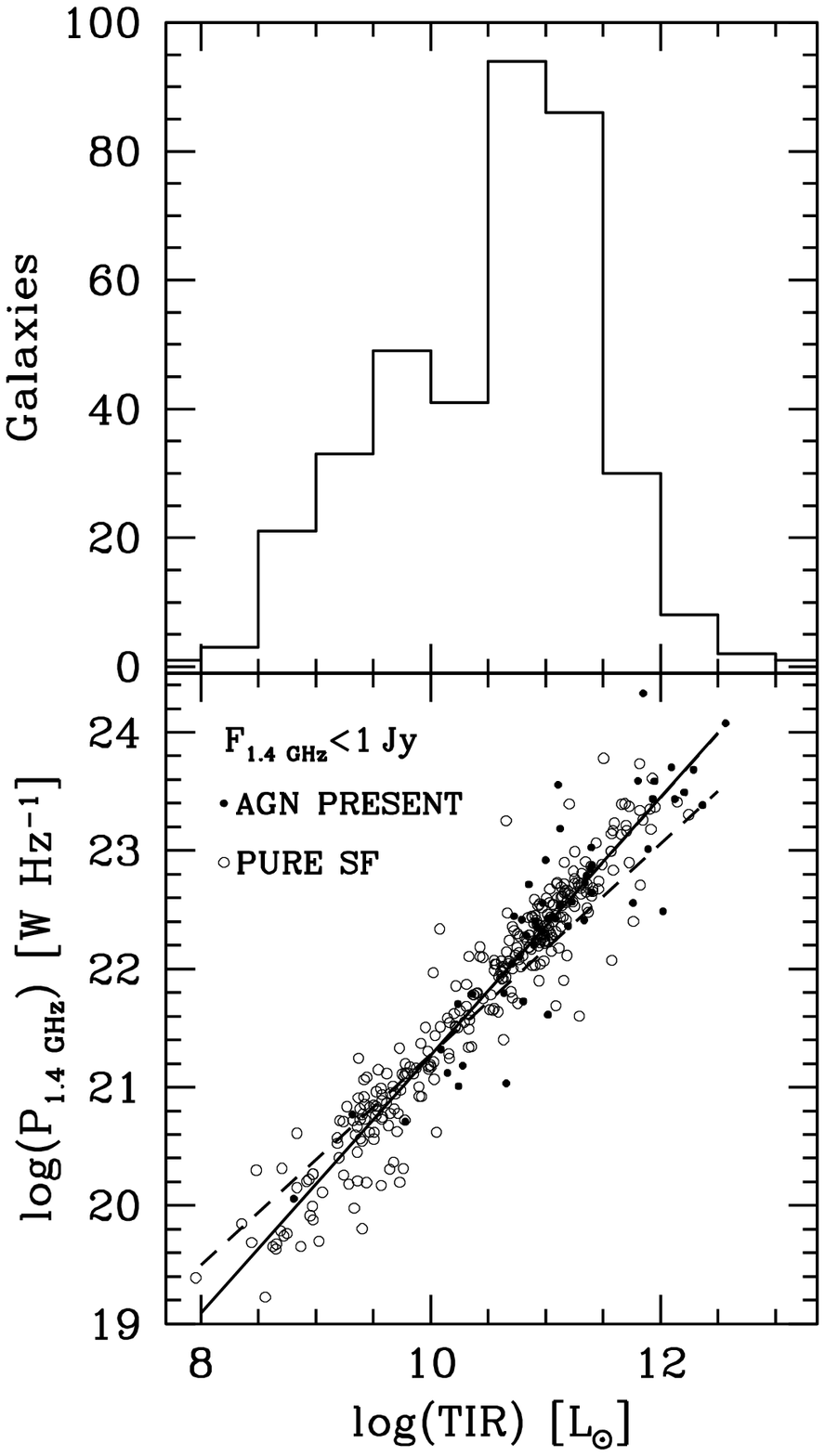}
\caption{
{\sl Upper panel:} Histogram of total infrared luminosities calculated
as described in the text.
{\sl Lower panel:} The FIR-radio correlation for 367 galaxies with
radio continuum fluxes below 1\,Jy, \ie, all but the two SFRS galaxies
(3C\,273 and OJ\,287) dominated by emission from an AGN at
radio wavelengths.
Open circles indicate the 52 sources (Table~\ref{tab:agn_frac})
for which at least one of the three AGN
criteria described in Sec.~\ref{sec:agn} is met.
Solid circles indicate sources not known to have any AGN contribution.
The line is an unweighted least-squares fit to the entire distribution.
The same fitting procedure, when performed only for the 315 `pure'
star-forming galaxies, yields a correlation indistinguishable
from that for all 367 galaxies on the scale of this plot.
The dashed line illustrates a slope of unity.
}
\label{fig:radiocorr}
\end{figure}

No matter how they are identified, the AGNs in the SFRS do not 
have a noticeably different $F_{60}/F_{100}$ ratio than the non-AGNs.
Figure~\ref{fig:agn_iras} is a far-infrared color-color diagram showing the 313 SFRS 
galaxies with either \IRAS\ 25\,$\mu$m or MIPS 24\,$\mu$m flux measurements,
including the 22 objects identified as AGNs by the Sanders \etal (1988) criterion.  
This suggests that for most galaxies in the sample, AGNs contribute 
relatively little of the far-infrared light.  Conversely, the far-infrared emission 
is an uncontaminated tracer of star formation.
This result is consistent with Mullaney \etal (2011), who found that the SEDs
of AGNs drop rapidly at wavelengths longer than 40\,$\mu$m.  It is also consistent with
the outcome of Netzer \etal (2007), who found that the far-infrared properties of QSOs is
almost entirely due to star formation.
This view is confirmed by an examination of the FIR-radio correlation 
(\eg, Condon \etal 1992, Helou \etal 1985), the fact 
that the non-thermal radio emission and the thermal far-infrared emission correlate
tightly over at least four orders of magnitude in far-infrared luminosity.
Fig.~\ref{fig:radiocorr} shows the FIR-radio correlation 
constructed using {\sl total} infrared flux estimates
$F(TIR)=2.403\nu f_{\nu}(25\,\mu {\rm m}) - 0.2454\nu f_{\nu}(60\,\mu {\rm m}) + 1.347\nu f_{\nu}(100\,\mu {\rm m})$
in the usual way (Dale \& Helou 2002) except that we employ MIPS 24\,$\mu$m fluxes
where available instead of the less precise \IRAS\ 25\,$\mu$m fluxes. 
The solid line in the lower panel of Fig.~\ref{fig:radiocorr} is an unweighted 
least-squares fit to all SFRS galaxies (except 3C\,273 and OJ\,287, which
have extremely high radio continuum luminosities arising from dominant central AGNs).
If all galaxies flagged as hosting AGNs
by at least one of the three criteria above are excluded from the fit, the slope
changes very little (to 1.09 from 1.086) and the scatter about the fit does 
likewise (decreasing to 0.29 dex from 0.31 dex).  Thus, if the two dominant AGNs
are excluded from the fit, the FIR-radio correlation is not significantly 
affected by the presence of AGNs in 52/367 (roughly one of every seven galaxies).
This underlines the utility of the thermal far-IR as an optimal probe of SFR.

The FIR-radio correlation obtained for the SFRS sample is higher than unity but
nonetheless shallower than reported by Devereux \& Eales (1989), roughly 1.1 
instead of 1.28.  This could be due to selection effects: Devereux \& Eales (1989) 
employed an optical selection that sampled luminosities only up to $\sim10^{11}$\,\Lsun.  
Yun, Reddy, \& Condon (2001) find a slope much closer to unity for
their sample of bright infrared-selected galaxies, a result later confirmed
by Bell (2003).

\tablenum{10}
\begin{deluxetable*}{rrccc}
\tabletypesize{\scriptsize}
\tablecaption{AGN Detections\label{tab:agn_frac}}
\tablewidth{0pt}
\tablehead{
\colhead{Number}     & \colhead{Name} & \multicolumn{3}{c}{Detection Method} \\
\colhead{}           & \colhead{}     & \colhead{Emission Lines}  & \colhead{$F_{24}/F_{60}$} & \colhead{3.6--8.0\,$\mu$m Colors}
}
\startdata
1   & IC\,486            &  Y       &     Y      & \nodata \\
7   & IRAS\,08072+1847   & \nodata  &     Y      & \nodata \\
30  & OJ\,287            &  Y       &     Y      &    Y    \\
32  & IRAS\,08550+3908   &  Y       &  \nodata   &    -    \\
36  & IRAS\,08572+3915SW & \nodata  &  \nodata   &    Y    \\
55  & NGC\,2893          & \nodata  &     Y      &    -    \\
57  & CGCG\,238-066      &  Y       &     Y      &    -    \\
64  & CGCG\,182-010      & \nodata  &     Y      &    -    \\
76  & IC\,2551           & \nodata  &     Y      &    -    \\
79  & IRAS\,10120+1653   &  Y       &  \nodata   & \nodata \\
91  & UGC\,5713          & \nodata  &     Y      &    Y    \\
100 & UGC\,5881          &  Y       &  \nodata   & \nodata \\
107 & CGCG\,95-055       &  Y       &  \nodata   & \nodata \\
109 & UGC\,6074          & \nodata  &     Y      & \nodata \\
118 & IC\,2637           &  Y       &  \nodata   & \nodata \\
121 & IRAS\,11167+5351   & \nodata  &  \nodata   &    Y    \\
122 & NGC\,3633          & \nodata  &     Y      & \nodata \\
139 & NGC\,3758          & \nodata  &  \nodata   &    Y    \\
169 & UGC\,7016          &  Y       &  \nodata   & \nodata \\
189 & NGC\,4253          & \nodata  &     Y      &    Y    \\
194 & NGC\,4385          & \nodata  &     Y      & \nodata \\
198 & NGC\,4418          & \nodata  &     Y      &    Y    \\
204 & 3C\,273            & \nodata  &     Y      &    Y    \\
233 & MCG\,8-23-097      &  Y       &  \nodata   & \nodata \\
239 & UGC\,8058          & \nodata  &     Y      &    Y    \\
244 & NGC\,4922          & \nodata  &  \nodata   &    Y    \\
248 & UGC\,8269          &  Y       &  \nodata   & \nodata \\
253 & IRAS\,13144+4508   &  Y       &     Y      & \nodata \\
259 & NGC\,5104          &  Y       &  \nodata   & \nodata \\
263 & IRAS\,13218+0552   & \nodata  &  \nodata   &    Y    \\
270 & IRAS\,13349+2438   & \nodata  &     Y      & \nodata \\
276 & MK\,268            &  Y       &  \nodata   &    Y    \\
277 & NGC\,5278          &  Y       &  \nodata   & \nodata \\
278 & NGC\,5273          &  Y       &  \nodata   & \nodata \\
281 & UGC\,8696          &  Y       &  \nodata   &    Y    \\
283 & MK\,796            & \nodata  &  \nodata   &    Y    \\
284 & IRAS\,13446+1121   &  Y       &  \nodata   &    Y    \\
286 & NGC\,5313          &  Y       &  \nodata   & \nodata \\
288 & NGC\,5347          &  Y       &     Y      &    -    \\
292 & UGC\,8850          & \nodata  &     Y      &    Y    \\
296 & NGC\,5403          &  Y       &  \nodata   & \nodata \\
303 & CGCG\,074-129      &  Y       &     Y      &    Y    \\
309 & IC\,4395           &  Y       &  \nodata   &    -    \\
322 & UGC\,9412          & \nodata  &     Y      &    Y    \\
329 & IRAS\,14538+1730   &  Y       &  \nodata   & \nodata \\
332 & UGC\,9639          &  Y       &  \nodata   & \nodata \\
339 & IC\,4553           &  Y       &  \nodata   &    -    \\
341 & IC\,4567           &  Y       &  \nodata   & \nodata \\
343 & NGC\,5975          &  Y       &  \nodata   & \nodata \\
350 & UGC\,10120         & \nodata  &     Y      &    Y    \\
352 & NGC\,6040          &  Y       &  \nodata   & \nodata \\
358 & IRAS\,16150+2233   &  Y       &  \nodata   & \nodata \\
\tableline
    & Total &  30      &    22      &    19   \\
    & Total AGNs (all methods) &    &    52      &         \\
\enddata
\end{deluxetable*}

In summary, the three different AGN detection criteria identify three 
different AGN subsamples comprising from 5 to 10\% of the SFRS 
study sample.  There is only modest overlap among the three criteria
(Table~\ref{tab:agn_frac}), which in hindsight validates the use of multiple
tracers of AGN activity.  AGNs are more prevalent at higher luminosities,
despite the fact that they contribute little to the total far-infrared luminosities
of star-forming galaxies.  AGNs were detected in $\sim$15\% of the 
sample (52/369 galaxies) when all three detection methods were combined.
Taken as a whole, these results suggest that 
the far-infrared selection described above (Sec~\ref{sec:selection}) 
has successfully yielded a sample with bolometric luminosities
dominated by star formation, and there is no compelling
evidence that the AGNs significantly contaminate the far-infrared SEDS.

\subsection{Pending Observing Campaigns}
\label{ssec:pending_obs}

There are several ongoing observing campaigns for which significant data have
been acquired but which are not yet complete.  They are described below.

\subsubsection{Longslit Optical Spectroscopy with FAST}
\label{sec:fast}

Visible spectra from SDSS are available only for a fraction of the SFRS galaxies 
(Abazajian \etal 2009).  Moreover, the SDSS spectra were obtained through 
fibers centered on the nuclei and so do not uniformly measure or constrain the 
excitation conditions within the galaxy disks.  We are therefore reobserving 
SDSS galaxies with a long-slit spectrograph in order to: 1.) build a consistent set 
of disk+nuclear spectra with better sky subtraction, 2.) take advantage of the spatial 
dimension to implement accurate starlight subtraction, and 3.) measure rotation curves.

The observations are currently being carried out at the Fred Lawrence Whipple Observatory 
1.5\,m telescope at Mt. Hopkins with FAST.
FAST is being used with a 2\arcsec\ wide slit positioned along the major axis of the galaxies.  
For galaxies which extend beyond the slit ends we obtain either a second 
spectrum along the minor axis of the galaxy, or a second spectrum at a sky position 
away from the galaxy.   The spectra are obtained with a 600\,l/mm grating in two tilt 
positions in order to cover the blue ($3700-5700$\,\AA) and the red ($5500-7500$\,\AA) 
part of the spectrum with resolutions of 2.2\,\AA\ and 2.7\,\AA\, respectively.  
Throughout the modest redshift range occupied by our sample, this setup covers the 
standard diagnostic emission lines in the optical band: 
[\ion{O}{2}]$\lambda3727$\,\AA; H$\beta$;  [\ion{O}{3}]$\lambda\lambda4959,5007$\,\AA;   
[\ion{O}{1}]$\lambda6300$\,\AA; [\ion{N}{2}]$\lambda\lambda6548,6583$\,\AA; H$\alpha$; 
and [\ion{S}{2}]$\lambda\lambda6716,6731$\,\AA.  The exposure time for each nucleus 
has been chosen in order to achieve a signal-to-noise ratio ($S/N$) of at least 
40 at the H$\alpha$ line, based on the available photometry from SDSS.

\subsubsection{PAIRITEL Near-IR Imaging}
\label{ssec:pairitel}

As described in Sec.~\ref{sec:selection}, 2MASS $K_{20}$ magnitudes were used as 
a stellar mass proxy to define the SFRS sample because of the well-characterized 
and uniform data quality and the full-sky coverage.  However, a campaign has been 
initiated to replace the 2MASS photometry with significantly deeper near-IR observations 
in the same three bands, because the 2MASS observations are somewhat shallow.  
They are not optimal for detecting the faint outskirts of even nearby galaxies, 
where relatively high sky 
backgrounds obscure the outermost, low surface-brightness features.  This leads to
a systematic downward bias in the 2MASS photometry that is at present not fully characterized.
Kirby \etal (2008) found the discrepancy to be highly variable, ranging up to 2.5\,mag
in extreme cases.  Karachentsev \etal (2002) found that the global 2MASS XSC photometry 
implied unphysical colors in some cases.  

Deeper near-infrared observations will offer several advantages.  They will greatly
reduce both the measurement uncertainties and the bias in the existing photometry.  They will
better characterize the true extents and morphologies of the SFRS galaxies.  They will
also be a much better match to the existing \SS/IRAC data, which benefit from the very low 
backgrounds available from space.  This will facilitate accurate K-corrections 
and stellar mass estimates planned for subsequent papers as well as reliable 
bulge/disk decompositions (because a greater extent and dynamic range in disk surface
brightness is sampled).

Motivated by these considerations, the SFRS team sought and was awarded time for near-infrared
imaging in 2009, 2010, and 2011 with the Peterson Automated Infrared Imaging Telescope (PAIRITEL; 
Bloom \etal 2006),  the same telescope originally used for 2MASS.  Useful $JHK_s$ imaging 
(\ie, flattened, background-subtracted, and astrometrically correct mosaics encompassing 
all the emission from the targets) has been acquired for 259 SFRS galaxies to date.
Exposure times were typically 20\,min on-source, but in some cases the exposures
were shorter because of circumstances related to weather or instrumentation.
The PAIRITEL observations typically reach two magnitudes deeper than 2MASS.  Based on a 
preliminary analysis of the $K_s$ imaging data reduced to date, it appears that 
the total magnitudes listed in Table~\ref{tab-phot1} understate the galaxies' true
output by about 0.3\,mag (P. Bonfini, 2011facility, private communication).
The goal is to complete the deep $JHK_s$ coverage in the
2012a observing semester, after which the imaging and photometry will be presented together
with a structural decomposition of the SFR galaxies (P.~Bonfini et al., 2012, in preparation).

\subsubsection{Far-Infrared Detections by AKARI}
\label{ssec:akari}

\Akari\ is a cryogenic space-based infrared telescope facility launched in 2006.  
Because it uses infrared detectors with pixels smaller (0\farcm5--0\farcm9; Jeong \etal 2007) 
than those used by \IRAS, \Akari\ offers spatial resolution superior 
to that of \IRAS\ despite having a primary mirror only slightly larger. 
For example, the Far-Infrared Surveyor (FIS) point spread functions have FWHMs 
ranging from 40\arcsec\ to 60\arcsec\ depending on the band (Kawada \etal 2007).
\Akari\ therefore offers a way to improve upon \IRAS\ photometry and likewise improve 
the fidelity of the far-IR SFRs inferred for our sample
in cases of high foreground gradients arising from Galactic cirrus and/or confused \IRAS\ sources,
because \Akari\ has the potential to resolve out such contributions to the 
\IRAS\ flux measurements.

Two all-sky surveys were carried out by the \Akari\ mission, covering $>$94\% of the sky more than 
twice (Murakami \etal 2007).  The central wavelengths of the six survey bands are 9 and 
18\,$\mu$m with the Infrared Camera (IRC; Ishihara \etal 2010) and 65, 90, 140, and 
160\,$\mu$m with the FIS (Kawada \etal 2007).  
All SFRS galaxies were detected in both the FIS and IRC all-sky surveys, which reached
detection limits of 0.21\,Jy at the 80\% completeness level at 18\,$\mu$m and 
a 5$\sigma$ detection limit of 0.55\,Jy at 90\,$\mu$m.  The 90\,$\mu$m band is
the most sensitive of the four FIS bands and covers a bandpass very similar to
the \IRAS\ 100\,$\mu$m band.  A subsequent paper (H. Kaneda et al., in preparation)
is planned to refine the far-IR SFR estimates with the full suite of \Akari\ photometry.

\section{Conclusion}
\label{sec:conclusion}

The Star Formation Reference Survey, by virtue of its reliance on a restricted 
but representative far-IR selection, ensures the capability to study 
{\sl obscured} star formation in all of its varied manifestations in
the local Universe.  Its panchromatic resources, spanning UV to radio
wavelengths and featuring 100\% complete photometry in the visible to mid-infrared regimes
($ugrizJHK_s$ and four IRAC bands) provides an opportunity to quantitatively 
assess the degree to which far-IR emission reflects total (not just obscured) 
SFR.  This comprehensive collection of the most widely-used SFR indicators 
will furnish an invaluable resource for the interpretation of more distant
galaxies, \ie, galaxies for which some or even nearly all such 
SFR metrics are inaccessible because of their relative faintness, so that 
a context exists in which to better understand the limited data available.
Because the SFRS is fully representative of star-forming galaxies in the local Universe, 
it is an optimal benchmark for understanding star formation
in the distant cosmos.

The interplay of various star formation and AGN indicators will be explored 
in future papers.  Paper~2 will examine the relationship of UV SFR indicators 
to those obtained in the other bands.
P. Bonfini \etal (2012, in preparation) will describe the methods whereby relatively
faint AGNs in the sample are identified using structural decomposition.
Y.-N. Zhu \etal (2012, in preparation) will examine a suite of narrow-band
\halpha\ imaging and compare the global \halpha-derived SFRs to those
in the other bands compiled for the SFRS.  All the related photometry 
will be made public to facilitate investigations by others in the community.

\acknowledgments

The authors gratefully acknowledge the assistance of R. Brent Tully, who generously 
supplied quality distances for many of our sample galaxies.  
This research has made use of the NASA/IPAC Extragalactic Database (NED) 
which is operated by the Jet Propulsion Laboratory, California Institute 
of Technology, under contract with the National Aeronautics and Space Administration.
This work is based in part on observations made with the Spitzer Space Telescope, 
which is operated by the Jet Propulsion Laboratory, California Institute of 
Technology under a contract with NASA. Support for this work was provided by NASA. 
The National Radio Astronomy Observatory 
is a facility of the National Science Foundation operated under cooperative 
agreement by Associated Universities, Inc.  
H. Wu is supported by NSFC grants 
10833006, 10773014, and the 973 Program grant 2007CB815406.  P. Barmby 
acknowledges research support through a Discovery Grant from the
Natural Sciences and Engineering Research Council of Canada.  S. Mahajan 
acknowledges support from a Smithsonian Institution Endowment Grant.
A. Zezas and P. Bonfini acknowledge support from Marie-Curie IRG grant 
224878 and European Union grant 206469.
H. Smith acknowledges partial support from NASA grants NNX07AH49G and NNX10AD83G.



{\it Facilities:} \facility{VLA}, \facility{\SS (IRAC, MIPS)}, \facility{\GG}, 
\facility{2MASS}, \facility{SDSS}, \facility{NAOC}.



\clearpage
\LongTables
\tablenum{2}
. Make sure there is at least one \tablenotemark
\tablecomments{Morphological T types and B-band Milky Way extinction (in magnitudes, estimated from
the 100\,$\mu$m dust maps) were taken from the PSCz catalog (Saunders \etal 2000), except as noted.
Axial ratios were measured directly from the IRAC mosaics using SExtractor and are estimated to be
accurate to 0.05 (1$\sigma$).
$L_{60}$ is a monochromatic 60\,$\mu$m luminosity calculated for purposes of sample selection.  The more
useful L(TIR) is given in Table~\ref{tab:fir}. 
Weights are the ratio of the total number of PSCz galaxies within a given bin to the number of SFRS
sample galaxies taken from that same bin, subject to the area restrictions noted in Sec.~2.
}

\tablenotetext{a}{Coordinates are based on IRAC imaging.}
\tablenotetext{b}{Quality distances from R. B. Tully (priv. comm.) are given
with 1$\sigma$ error estimates.  All other distances are derived from the PSCz velocities
after correcting for local bulk flow as described in Sec.~\ref{sec-velocities}. }
\end{deluxetable}

\clearpage

\tablenum{4}
\begin{deluxetable}{rrcccc}
\tabletypesize{\scriptsize}
\tablecaption{\SS/IRAC Photometry for SFRS Galaxies\label{tab-irac1}}
\tablewidth{0pt}
\tablehead{
\colhead{SFRS} & \colhead{Name} & \colhead{$3.6\,\mu$m} & \colhead{$4.5\,\mu$m} & \colhead{$5.8\,\mu$m} & \colhead{$8.0\,\mu$m} \\
}
\startdata
  1 &           IC\,486 & 12.58         & 12.71         & 12.37         & 11.81         \\
  2 &          IC\,2217 & 13.06         & 13.44         & 11.95         & 10.78         \\
  3 &         NGC\,2500 & 11.78         & 12.25         & 11.48         & 10.95         \\
  4 &         NGC\,2512 & 12.25         & 12.64         & 11.67         & 10.63         \\
  5 &     MCG\,6-18-009 & 13.14         & 13.50         & 12.80         & 11.51         \\
  6 &          MK\,1212 & 13.65         & 13.90         & 12.85         & 11.37         \\
  7 &  IRAS\,08072+1847 & 13.81         & 13.35         & 12.11         & 10.94         \\
  8 &         NGC\,2532 &     \nodata   & 12.22         &    \nodata    &  9.97         \\
\enddata
\tablecomments{Table~\ref{tab-irac1} is published in its entirety in the
electronic edition of {\it Publications of the Astronomical Society of the Pacific}.
All magnitudes are expressed on the AB system.  Where uncertainties are not given,
they are dominated by the uncertainty in the cryogenic IRAC absolute flux calibration
(3\%, 1$\sigma$).
}
\end{deluxetable}

\clearpage

\tablenum{5}
\begin{deluxetable}{rrcrrcl}
\tabletypesize{\scriptsize}
\tablecaption{1.4\,GHz Flux Measurements for SFRS Galaxies\label{tab-radio}}
\tablewidth{0pt}
\tablehead{
\colhead{SFRS} & \colhead{Name} & \colhead{Position\tablenotemark{a}} & \colhead{Offset\tablenotemark{b}} & \colhead{F(1.4\,GHz)} & \colhead{References\tablenotemark{c}} & \colhead{Remarks\tablenotemark{d}} \\
\colhead{}       & \colhead{} & \colhead{(J2000)} & \colhead{(arcsec)} & \colhead{(mJy)} & & \\
}
\startdata
  1 &           IC\,486 & 08:00:20.98 +26:36:50.2 &       1.5 &      10.2$\pm$0.5 & 1 &    \\ 
  2 &          IC\,2217 & 08:00:49.89 +27:30:01.9 &       2.2 &      18.8$\pm$0.7 & 1 &    \\ 
  3 &         NGC\,2500 & 08:01:52.41 +50:44:25.6 &      14.0 &      14.7$\pm$3.2 & 1 &    E*\\ 
  4 &         NGC\,2512 & 08:03:07.86 +23:23:31.6 &       1.1 &      18.6$\pm$0.7 & 1 &    \\ 
  5 &     MCG\,6-18-009 & 08:03:28.82 +33:27:44.4 &       1.5 &      17.9$\pm$0.7 & 1 &    \\ 
  6 &          MK\,1212 & 08:07:05.61 +27:07:33.8 &       1.2 &      12.4$\pm$0.6 & 1 &    \\ 
  7 &  IRAS\,08072+1847 & 08:10:07.18 +18:38:17.8 &       2.4 &       6.8$\pm$0.4 & 1 &    \\ 
  8 &         NGC\,2532 & 08:10:15.43 +33:57:25.9 &       3.9 &      46.5$\pm$2.1 & 1 &    \\
\enddata
\end{deluxetable}

\clearpage

\tablenum{7}
\begin{deluxetable}{rrrrrrrrr}
\tabletypesize{\scriptsize}
\tablecaption{Far-Infrared Photometry for SFRS Galaxies\label{tab:fir}}
\tablewidth{0pt}
\tablehead{
\colhead{SFRS} & \colhead{Name} & \colhead{L(TIR)} & \colhead{MIPS F(24)\tablenotemark{a}} & \colhead{IRAS F(60)} & \colhead{IRAS F(100)} &
\colhead{Planck F(350)} & \colhead{Planck F(550)} & \colhead{Planck (F850)} \\
\colhead{}       & \colhead{} & \colhead{(\Lsun)} & \colhead{(Jy)} & \colhead{(Jy)} & \colhead{(Jy)} & \colhead{(Jy)} & \colhead{(Jy)} & \colhead{(Jy)} \\
}
\startdata
  1 &                      IC\,486 &   10.90 & 0.44\tablenotemark{b} &   0.99 &               $<$1.51 &           \nodata &           \nodata &           \nodata \\ 
  2 &                     IC\,2217 &   10.74 &               $<$0.32 &   2.46 &                  4.81 &           \nodata &           \nodata &           \nodata \\ 
  3 &                    NGC\,2500 &    9.34 &                  0.22 &   2.81 &                  5.75 &     3.02$\pm$0.14 &           \nodata &           \nodata \\ 
  4 &                    NGC\,2512 &   10.88 & 0.66\tablenotemark{b} &   3.80 &                  7.31 &           \nodata &           \nodata &           \nodata \\ 
  5 &                MCG\,6-18-009 &   11.25 &               $<$0.27 &   1.65 &                  3.09 &           \nodata &           \nodata &           \nodata \\ 
  6 &                     MK\,1212 &   11.35 &               $<$0.33 &   1.87 &                  3.36 &           \nodata &           \nodata &           \nodata \\ 
  7 &             IRAS\,08072+1847 &   10.72 & 0.74\tablenotemark{b} &   2.79 &                  3.08 &           \nodata &           \nodata &           \nodata \\ 
  8 &                    NGC\,2532 &   11.09 &                  0.63 &   3.61 &                 10.34 &     4.02$\pm$0.21 &     1.43$\pm$0.08 &           \nodata \\
\enddata
\tablecomments{}
\tablenotetext{a}{The MIPS 24\,$\mu$m photometry has been aperture corrected using the correction
factors listed in Table~\ref{tab-mips}.  The uncertainties in total 24\,$\mu$m flux densities are dominated
by the uncertainty in the absolute calibration, which is estimated as 4--8\% according to the MIPS
Instrument Handbook, version 2.  Where upper limits are given, the photometry is from IRAS with
quality flag = 1.}
\tablenotetext{b}{IRAS flux density (MIPS not available).  Typical uncertainties are 10--20\%.}
\tablenotetext{c}{OJ\,287 is highly variable.  Flux densities here are from \IRAS\ survey coadds 
\citep{Impey1988}, but many more measurements exist in the literature.}
\tablenotetext{d}{IRAS flux densities taken from Rice \etal 1988.}
\tablenotetext{e}{Measurement taken from Dale \etal 2005.}
\end{deluxetable}

\clearpage

\tablenum{8}
\begin{deluxetable}{rrc}
\tabletypesize{\scriptsize}
\tablecaption{Spitzer/MIPS Aperture Correction Factors for SFRS Galaxies\label{tab-mips}}
\tablewidth{0pt}
\tablehead{
\colhead{SFRS} & \colhead{Name} & \colhead{Factor} \\ 
}
\startdata
 24  & NGC\,2623 & 1.01 \\ 
 35  & NGC\,2719 & 1.01 \\ 
 36  & IRAS\,08572+3915 & 1.04 \\ 
 52  & NGC\,2854 & 1.05 \\ 
 60  & MCG\,8-18-013 & 1.05 \\ 
 70  & NGC\,3049 & 1.01 \\ 
 85  & UGC\,5644 & 1.07 \\ 
 90  & NGC\,3265 & 1.03 \\
\enddata
\tablecomments{MIPS 24\,$\mu$m aperture correction factors based
on Version 2 of the MIPS Instrument Handbook, page 94.  These factors
have been applied to the photometry presented in Table~\ref{tab:fir}.
When not given, the correction factor was taken to be unity.
}
\end{deluxetable}

\clearpage

\tablenum{9}
\begin{deluxetable}{ccccccccc}
\tabletypesize{\scriptsize}
\tablecaption{Visible and Near-Infrared Photometry for SFRS Galaxies\label{tab-phot1}}
\tablewidth{0pt}
\tablehead{
\colhead{SFRS} & \multicolumn{5}{c}{SDSS}                                                                  & \multicolumn{3}{c}{2MASS} \\
\colhead{}  & \colhead{$u$}    & \colhead{$g$}    & \colhead{$r$}    & \colhead{$i$}    & \colhead{$z$}    & \colhead{$J$}    & \colhead{$H$}  & \colhead{$K_s$}\\
}
\startdata
  1 & 15.838$\pm$0.015 & 14.095$\pm$0.002 & 13.370$\pm$0.004 & 13.009$\pm$0.002 & 12.807$\pm$0.005 &	12.38$\pm$0.02 & 12.18$\pm$0.02 & 12.34$\pm$0.03 \\
  2 & 15.387$\pm$0.025 & 14.100$\pm$0.010 & 13.683$\pm$0.018 & 13.407$\pm$0.023 & 13.289$\pm$0.023 &	12.70$\pm$0.02 & 12.54$\pm$0.03 & 12.74$\pm$0.04 \\
  3 & 15.796$\pm$0.049 & 13.236$\pm$0.008 & 12.769$\pm$0.006 & 12.985$\pm$0.017 & 13.718$\pm$0.032 &	11.66$\pm$0.02 & 11.52$\pm$0.03 & 11.12$\pm$0.07 \\
  4 & 15.413$\pm$0.018 & 13.567$\pm$0.004 & 12.873$\pm$0.003 & 12.531$\pm$0.003 & 12.321$\pm$0.005 &	11.73$\pm$0.02 & 11.47$\pm$0.03 & 11.66$\pm$0.03 \\
  5 & 15.966$\pm$0.019 & 14.575$\pm$0.003 & 13.883$\pm$0.004 & 13.486$\pm$0.003 & 13.239$\pm$0.005 &	12.67$\pm$0.03 & 12.49$\pm$0.03 & 12.60$\pm$0.04 \\
  6 & 16.531$\pm$0.015 & 15.256$\pm$0.027 & 14.641$\pm$0.033 & 14.277$\pm$0.025 & 14.062$\pm$0.025 &	13.46$\pm$0.03 & 13.35$\pm$0.03 & 13.43$\pm$0.04 \\
  7 & 18.184$\pm$0.043 & 16.437$\pm$0.005 & 15.549$\pm$0.003 & 15.117$\pm$0.004 & 14.790$\pm$0.008 &	14.14$\pm$0.04 & 14.04$\pm$0.06 & 14.03$\pm$0.06 \\
  8 & 15.073$\pm$0.025 & 13.432$\pm$0.007 & 12.865$\pm$0.004 & 12.500$\pm$0.005 & 12.571$\pm$0.009 &	11.54$\pm$0.02 & 11.36$\pm$0.03 & 11.45$\pm$0.04 \\
\enddata
\tablecomments{SDSS DR7 Petrosian magnitudes and 2MASS total magnitudes for the SFRS
galaxies.  Where the SDSS photometry is unreliable due to shredding, no values are given.
Uncertainties are taken from the 2MASS extended source catalog and reflect measurement
errors only.  Issues pertaining to systematic errors in the 2MASS data are discussed
in Sec.~\ref{ssec:pairitel}
Uncertainties are taken from the 2MASS extended source catalog and reflect measurement
errors only.  Issues pertaining to systematic errors in the 2MASS data are discussed
in Sec.~\ref{ssec:pairitel}
All magnitudes are expressed on the AB system.  }
\tablenotetext{a}{SDSS $u$ photometry was replaced by an estimate based
on applying the Jester \etal (2005) filter transformations to the global $BVRI$ photometry 
from Hyperleda.}
\tablenotetext{b}{Available SDSS $ugriz$ photometry corresponds only to a portion of this source.}
\tablenotetext{c}{SDSS $z$ photometry was replaced by an estimate based
on applying the Jester \etal (2005) filter transformations to the global $BVRI$ photometry 
from Hyperleda.}
\tablenotetext{d}{Available SDSS $z$ photometry corresponds only to a portion of this source.}
\tablenotetext{e}{SDSS $ugriz$ photometry was replaced by an estimate based
on applying the Jester \etal (2005) filter transformations to the global $BVRI$ photometry 
from Hyperleda.}
\tablenotetext{f}{SDSS $gr$ photometry was replaced by an estimate based
on applying the Jester \etal (2005) filter transformations to the global $BVRI$ photometry 
from Hyperleda.}
\tablenotetext{g}{SDSS $ugr$ photometry was replaced by an estimate based
on applying the Jester \etal (2005) filter transformations to the global $BVRI$ photometry 
from Hyperleda.}
\end{deluxetable}

\clearpage

\end{document}